\documentclass[times,final]{elsarticle}

\usepackage{amsmath} 
\usepackage{amssymb}

\usepackage{jcomp}

\usepackage{multicol}
\usepackage[linesnumbered,ruled,vlined]{algorithm2e}
\usepackage{amsmath}
\usepackage{eqparbox}
\usepackage{jcomp}
\usepackage{hyperref}
\usepackage{framed,multirow}

\hypersetup{
    colorlinks = true,
    urlcolor   = blue,
    citecolor  = red,
}

\usepackage{amssymb}
\usepackage{latexsym}
\usepackage{bm}
\usepackage{subfigure}
\usepackage{booktabs}
\usepackage{url}
\usepackage{xcolor}
\usepackage{float}
\definecolor{newcolor}{rgb}{.8,.349,.1}

\definecolor{Reviewer1Color}{RGB}{255, 0, 0}    
\definecolor{Reviewer2Color}{RGB}{255, 169, 0}    
\definecolor{Reviewer3Color}{RGB}{30, 144, 255}    

\usepackage{url}

\journal{Journal of Computational Physics}

\begin{document}

\verso{Z. H. Xue, J. Magnaudet, J. Zhang~\textit{et al.}}

\begin{frontmatter}

  \title{A sharp and conservative method for modeling interfacial flows with insoluble surfactants in the framework of a geometric volume-of-fluid approach}%

  \author[1]{Zhong-Han Xue}
  \author[2]{Jacques Magnaudet}
  \author[1]{Jie Zhang\corref{cor1}}
  \cortext[cor1]{Corresponding author:}
  \ead{j\_zhang@xjtu.edu.cn}

  \address[1]{ State Key Laboratory for Strength and Vibration of Mechanical Structures, School of Aerospace, Xi'an Jiaotong University, Xi'an 710049, China}
  \address[2]{ Institut de M\'ecanique des Fluides de Toulouse, University de Toulouse, CNRS, Toulouse 31400, France}

  \begin{abstract}
    Insoluble surfactants adsorbed at liquid-liquid or gas-liquid interfaces alter interfacial tension, leading to variations in the normal stress jump and generating tangential Marangoni stresses that can dramatically affect the flow dynamics. We develop a three-dimensional, sharp and conservative numerical method for modeling insoluble surfactant-laden interfacial flows within a volume-of-fluid framework. This method contrasts with diffusive transport algorithms commonly employed in the Eulerian framework. The proposed method preserves the zero-thickness property of the interface, ensures accurate calculation of the surfactant concentration, and robustly handles complex topological changes. The interface evolution is captured using a geometrical volume-of-fluid method, with surfactant mass sharply stored at the reconstructed interface. The advection term in the surfactant transport equation is discretized implicitly in conjunction with the geometrical advection of the volume fraction of one of the  fluids, thereby eliminating numerical inconsistencies arising from discrepancies between the actual and computed interface areas. Additionally, the diffusion term is discretized along the reconstructed interface, preventing artificial diffusion normal to the zero-thickness interface. Benchmark tests demonstrate that the proposed method achieves higher accuracy and faster convergence compared to existing diffusive approaches. Finally, we apply the method to investigate the interaction of a surfactant-laden rising bubble with a vertical wall, revealing a transition from near-wall bouncing to migration away from the wall as the surfactant concentration increases.
    \begin{keyword}
      insoluble surfactant; volume-of-fluid; sharp interface;
    \end{keyword}
  \end{abstract}

\end{frontmatter}

\section{Introduction}\label{sec:introduction}

In a variety of multiphase and free-surface flows, surfactants (surface-active agents), composed of a hydrophilic head and a hydrophobic tail, are transported in the bulk liquid and aggregate at the fluid-fluid interfaces, where they deeply influence the flow dynamics. Essentially, the presence of surfactants affects the flow characteristics through two distinct mechanisms. First, the interfacial tension, which depends on the surfactant concentration, is locally lowered, leading to a variation in the normal stress jump condition across the interface. Second, the nonuniform interfacial tension induces tangential shear stresses along the interface, resulting in the formation of a Marangoni force. These effects play a crucial role in a variety of natural phenomena and engineering applications, in domains such as oceanography, microfluidics, and coating technology \cite{lohse_surfactants_2023, weinstein_coating_2004, kovalchuk_review_2023}. It is important to note that surfactants can be either insoluble or soluble. Insoluble surfactants aggregate and transport along the zero-thickness interface, while soluble surfactants also exist in a dissolved form within the bulk liquid, and surfactant fluxes are exchanged with the interface through diffusion across a boundary layer. To understand the mechanisms governing the effects of surfactant on flow dynamics, it is essential to address the complex couplings between the fluid flow, the interface evolution, and the surfactant transport processes.

Experimentally, setting up well-controlled initial conditions and measuring interfacial quantities, such as the surfactant concentration, poses significant challenges, making detailed experimental investigations notoriously difficult, for instance in bubbly flows \cite{takagi_surfactant_2011,takemura_transverse_2003}. Numerical simulations offer a promising alternative to understand the coupled dynamics governing the fluid flow, interface evolution, and surfactant transport, thereby providing valuable insight into the basic mechanisms of contamination effects in two-phase flows. However, the development of accurate and robust numerical methods for modeling surfactant-laden flows remains a complex task. Specifically, insoluble surfactants are confined to the deforming and zero-thickness interface, necessitating the conservation of their interfacial mass. The accurate numerical representation of this zero-thickness transport is crucial for obtaining reliable predictions, especially over long time periods. The modeling of soluble surfactants introduces additional complexity, as accurately describing exchanges between the interfacial and bulk phases requires more advanced numerical schemes. This is why the present study focuses on insoluble surfactants, with the numerical methodology for soluble surfactants to be addressed in a subsequent paper.

Early numerical approaches to model surfactant-laden flows often relied on simplified models and did not fully address the coupled nature of the governing equations. In the context of Stokes flow, Stone and Leal \cite{stone_effects_1990} used a boundary integral method to investigate the effects of surfactants on drop deformation and breakup. Similarly, Yon and Pozrikidis \cite{yon_finite-volumeboundary-element_1998} proposed a hybrid finite volume/boundary element method to simulate the shear flow past a viscous drop. Assuming a spherical shape, Cuenot \textit{et al.} \cite{cuenot_effects_1997} solved the coupled Navier-Stokes and surfactant transport equations past a rising bubble to study the impact of slightly soluble surfactants on flow dynamics and drag. These approaches were limited by the lack of accurate and versatile numerical techniques capable of solving the fully coupled nonlinear equations with the temporal evolution of the fluid-fluid interface.

More recently, fully-coupled approaches have been developed to address the complexities of surfactant-laden flows involving deformable interfaces. These methods can generally be classified into two main categories, based on the approach they employ to discretize the transport equation of the surfactant. \textit{Lagrangian}-based methods allocate the surfactant concentration information directly to Lagrangian markers or elements. The front tracking method belongs to this category. Muradoglu and Tryggvason \cite{muradoglu_front-tracking_2008} introduced a finite difference/front tracking method to study three-dimensional flows involving soluble surfactants, further applying this approach to investigate the transient motion of buoyancy-driven bubbles \cite{tasoglu_effect_2008}. De Jesus \textit{et al.} \cite{de_jesus_3d_2015} developed a three-dimensional (3D) finite volume/front tracking method, making use of the Leibniz transport formula to express the time derivatives of integrals taken over moving and deformable interfaces. Their results were found to be consistent with experimental observations of deformed droplets contaminated by surfactants in shear flows. Similarly, Khati and Tornberg \cite{khatri_embedded_2014} employed the Segment Projection Method to transport surfactants along the interface. This method projects the surfactant concentration onto segments aligned with the coordinate system, allowing the surfactant transport equation to be solved along each segment. A notable advantage of these \textit{Lagrangian}-based methods lies in their ability to directly store the surfactant mass at the position of the Lagrangian markers or elements which move with the temporally evolving interface. This feature effectively circumvents the need for complex discretization schemes for the advection term in the surfactant transport equation, which must account for the stretching or compression of the interface. Consequently, these methods ensure that the sharpness of the interface is preserved throughout the simulation, even though it undergoes deformation. However, they have yet to demonstrate their efficiency in modeling highly deformed interfacial flows in 3D. Such problems are often characterized by complex topological changes, including coalescence and break-up, which can dramatically alter the dynamics of drop-laden or bubbly flows \cite{lu_multifluid_2019}.

The second family of approaches solves the interfacial surfactant transport within an Eulerian framework. Unlike Lagrangian methods, \textit{Eulerian}-based methods follow the evolution of the interfacial concentration of surfactants within discretized control volumes that are stationary with respect to the computational grid. While these methods are capable of tracking interface deformation and surfactant transport, they introduce additional challenges due to the need to employ schemes compatible with the zero-thickness interface. This results in an artificial smearing of the interface which makes the surfactants diffuse into an adjacent `spreading region'. Xu and Zhao \cite{xu_eulerian_2003} introduced a level set method to solve partial differential equations (PDEs) along a moving interface, representing the interface as a diffusive 3D tube (or 2D band) and discretizing the transport equation for insoluble surfactants. Xu \textit{et al.} \cite{xu_level-set_2018} later extended this method to accommodate soluble surfactants and applied it to study droplets interaction in 3D flows. Cleret de Langavant \textit{et al.} \cite{cleret_de_langavant_level-set_2017} further enhanced the method by incorporating quadtree/octree grids and a contact angle model to simulate phenomena such as the `tears of wine', with results exhibiting good agreement with experimental data. In recent years, the diffusive strategy has been integrated into other interface-capturing methods within the Eulerian framework, such as the diffusive interface method \cite{erik_teigen_diffuse-interface_2011}, the algebraic volume of fluid (VOF) method \cite{antritter_two-field_2024}, phase-field methods \cite{yamashita_conservative_2024}, a combination of both \cite{farsoiya_coupled_2024, jain_model_2024}, and lattice-Boltzmann methods \cite{ba_hybrid_2024}. These diffusive approaches circumvent the complexity associated with designing sharp numerical schemes for discretizing surfactant transport on a zero-thickness interface, making them numerically stable and relatively straightforward to implement within the Eulerian framework. However, these methods inherently fail to preserve the zero-thickness limit of the interface during simulations. This limitation leads to a reduced accuracy (typically of first order), and poor mass conservation properties \cite{cleret_de_langavant_level-set_2017}.

This state of the art suggests that there is significant potential in developing a sharp and conservative interface method for solving surfactant-related problems within the Eulerian framework. Such a method should resolve the fully coupled system of equations, ensuring the preservation of the zero-thickness property of the interface while also accommodating complex topological changes. However, despite this promise, \textit{Eulerian}-based sharp interface methods remain scarce due to the numerical challenges previously outlined. James and Lowengrub \cite{james_surfactant-conserving_2004} pioneered the modeling of surfactant-laden flows using a geometrical volume-of-fluid (VOF) method (limited to 2D axisymmetric configurations) in such a way that the interface sharpness is maintained. Their approach discretizes both surfactant mass and interface area geometrically, thereby avoiding the need to discretize interfacial surfactant transport accounting for interface stretching and compression. However, this strategy led to an inconsistency between the actual and reconstructed interface areas, as the interface area was primarily determined through geometric construction within the VOF framework. Since the tangential interfacial tension depends on the concentration gradient, which not only relies on the surfactant mass but also on the computed interface area, the accuracy of the method decreases over time, making it inappropriate for long-term simulations. Still in the framework of the geometrical VOF method, Alke and Bothe \cite{alke_3d_2009} proposed a 3D approach for modeling soluble surfactants, coupling the bulk and interfacial surfactant concentrations through an adsorption isotherm. This method is only suitable for scenarios involving fast surfactant sorption and is therefore limited to significantly soluble surfactants. As a result, no \textit{Eulerian}-based sharp interface method currently exists to model two-phase flows in the presence of insoluble surfactants, providing the central motivation to the present work.

In the present study, we propose a sharp and conservative numerical method for modeling 3D interfacial problems involving insoluble surfactants, where the fluid-fluid interface is tracked and reconstructed using the geometrical volume-of-fluid (VOF) approach. The main contribution of this work lies in addressing some long-standing numerical challenges inherent to \textit{Eulerian}-based methods. The key innovations of this method are as follows: First, insoluble surfactant mass and concentration are stored and transported on the reconstructed interface in a sharp manner, ensuring the preservation of the zero-thickness property of the interface. Second, the advection term in the surfactant transport equation is discretized implicitly alongside the geometrical advection of the VOF equation, establishing a consistent link between the computed and reconstructed interface areas, and avoiding numerical inconsistencies between the two quantities. Finally, the diffusion term is discretized along the geometrically reconstructed interfaces, ensuring that no artificial diffusion occurs in the direction normal to the zero-thickness interface. The paper is organized as follows: The physical problem and the numerical difficulties are described in \S~\ref{sec:physical-model}, with more details regarding currently available diffusive and sharp interface methods. The original numerical methods developed in the framework of the present study are presented in detail in \S~\ref{sec:two-phase-flow-solver} and \S~\ref{sec:sharp-scheme-surfactant-transport}. A series of benchmark test cases are presented in \S~\ref{sec:validation-cases}, where we demonstrate that the proposed approach outperforms diffusive methods in terms of accuracy. Finally, \S~\ref{sec:numerical-examples} showcases the capabilities of the present approach through a more challenging numerical example.

\section{Governing equations and numerical difficulties}
\label{sec:physical-model}
\begin{figure}
  \centering
  \includegraphics[width=\textwidth]{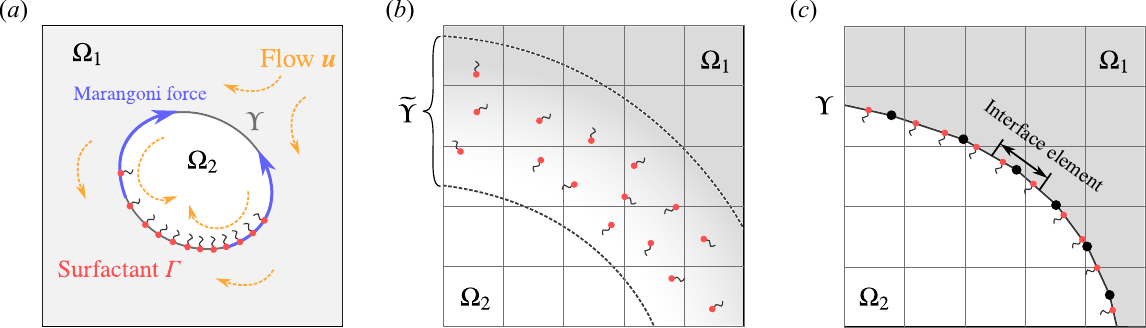}
    \vspace{-3mm}
  \caption{(\textit{a}) Schematic representation of the physical model. The computational domain contains two immiscible fluids, $\Omega_1$ and $\Omega_2$, separated by a deformable interface $\Upsilon$. 
  An insoluble surfactant (red points) lies at the interface with a surface concentration $\mathit{\Gamma}$. The surfactant reduces the interfacial tension $\sigma$ and makes it nonuniform, inducing Marangoni stresses (blue arrows) along the interface. (\textit{b}) Illustration of the \textit{Eulerian}-based diffuse interface method, where the surfactant is stored and transported within a finite-width interfacial region lying between the two phases. (\textit{c}) Illustration of the \textit{Lagrangian}-based sharp interface method, where the surfactant is stored and transported along Lagrangian zero-width interface elements. }
  \label{fig:schema}
\end{figure}

\subsection{Problem statement}\label{sec:problem-statement}

The physical model under consideration is illustrated in Fig.~\ref{fig:schema}(\textit{a}). The computational domain, denoted as $\Omega$, contains two immiscible fluid phases, phase 1 ($\Omega_1$) and phase 2 ($\Omega_2$), separated by a deformable interface $\Upsilon$. An insoluble surfactant is assumed to lie on the interface $\Upsilon$ and undergoes transport and diffusion along the interface. This surfactant is characterized by a local mass $M$ and a local surface concentration $\mathit{\Gamma}$, linked by the relation $\mathit{\Gamma} = M/A$, with $A$ representing the local interfacial length (in 2D) or area (in 3D) within a discretized control volume. In what follows, we will term $A$ the discretized interfacial area irrespective of the 2D or 3D nature of the flow configuration. The presence of the surfactant reduces the local value of the interfacial tension, generating spatial nonuniformities along the interface. These variations induce a Marangoni force (represented by the blue arrows in Fig.~\ref{fig:schema}(\textit{a})) which acts tangentially along the interface. Consequently, the dynamical consequences of the interfacial tension may be decomposed into two components. One acts in the direction normal to the interface and influences the normal stress balance, and thereby the shape of the interface. The second acts in directions tangential to the interface and is responsible for shear stresses that alter the fluid velocity $\bm{u}$ in the vicinity of the interface. The flow field, in turn, influences both the deformation of the interface and the transport of surfactants. This dynamic coupling results in a complex system involving fluid flow, surfactant transport, and the evolution of the interface.

\subsection{Governing equations}\label{sec:governing-equations}

The velocity field $\bm{u}$ throughout the fluid domain $\Omega=\Omega_1\cup\Upsilon\cup\Omega_2$ is governed by the `one-fluid' formulation of the Navier-Stokes equations, namely
\begin{equation}\label{eq:ns1}
  \rho\left(\frac{\partial \bm{u}}{\partial t} + (\bm{u} \cdot \nabla)\bm{u}\right) = -\nabla p + \nabla \cdot \left(\mu (\nabla\bm{u} + \nabla\bm{u}^T)\right) + \bm{F}, \qquad \forall\bm{x} \in \Omega,
\end{equation}
\begin{equation}\label{eq:ns2}
  \nabla \cdot \bm{u} = 0, \qquad \forall\bm{x} \in \Omega,
\end{equation}
where $p$ denotes the pressure, $\rho$ is the density, $\mu$ is the dynamic viscosity, and $\bm{F}$ represents the interfacial tension force acting on the interface. The specific form of $\bm{F}$ is discussed below.

In VOF approaches, Eqs. \eqref{eq:ns1}-\eqref{eq:ns2} are supplemented by a transport equation for the volume fraction $c$ which allows the two phases to be distinguished. Specifically, $c$ denotes the volume fraction of one phase, say phase 1, within an elementary control volume, i.e. a computational cell. Thus, the density is defined as $\rho = c \rho_1 + (1-c) \rho_2$ and a similar definition, $\mu = c \mu_1 + (1-c) \mu_2$, is used heuristically for the viscosity, with subscripts $1$ and $2$ referring to phases 1 and 2, respectively. The volume fraction $c$ evolves according to the transport equation
\begin{equation}\label{eq:vof-adv}
  \frac{\partial c}{\partial t} + (\bm{u} \cdot \nabla)c = 0, \qquad \forall\bm{x} \in \Omega .
\end{equation}

In the presence of a nonuniform interfacial tension, the stress jump condition across the interface reads \cite{pesci2018computational}
\begin{equation}\label{eq:jump-conditions}
  [\![p\mathbb{I} - \mu(\nabla\bm{u} + \nabla\bm{u}^T)]\!] \cdot \bm{n} = \sigma \kappa \bm{n} + \nabla_s \sigma, \qquad \forall\bm{x} \in \Upsilon,
\end{equation}
where the notation $[\![\cdot]\!]$ denotes the jump of a physical quantity across the interface from phase 1 to phase 2, $\bm{n}$ is the unit normal vector directed from phase 2 to phase 1, $\sigma$ is the interfacial tension which depends on the local surfactant concentration $\mathit{\Gamma}$, $\kappa = -\nabla \cdot \bm{n}$ is the interface mean curvature, and $\nabla_s = (\mathbb{I} - \bm{n} \bm{n}^T) \cdot \nabla$ is the surface gradient operator. Terms $\sigma \kappa \bm{n}$ and $\nabla_s \sigma$ represent the normal and tangential components of the stress jump, respectively.

Eq.~(\ref{eq:jump-conditions}) highlights the two key effects of surfactants on the interfacial stress jump. First, surfactants modify the interfacial tension $\sigma$ along the interface, which alters the normal stress jump. Second, the tangential gradient of the interfacial tension, $\nabla_s \sigma$, introduces a non-zero tangential stress jump. In Eq.~(\ref{eq:ns1}), these effects are incorporated in the body force $\bm{F}$ in the form
\begin{equation}\label{eq:body-force}
  {\bm{F}} = (\sigma \kappa  {\bm{n}}+ \nabla_s \sigma) \delta_s, \qquad \forall \bm{x} \in \Omega,
\end{equation}
where $\delta_s$ is the Dirac delta function centered on the interface $\Upsilon$. In this formulation, $\sigma \kappa {\bm{n}}\delta_s$ represents the Laplace pressure jump created by the interfacial tension, whereas $\nabla_s \sigma \delta_s$ represents the tangential Marangoni force induced by its variations along the interface.

The relation between the interfacial tension and the local surfactant concentration $\mathit{\Gamma}$ is usually modeled through a nonlinear equation of state (EOS). A commonly used model valid at low surfactant concentrations is the Langmuir adsorption isotherm \cite{levich1963physicochemical}, which reads
\begin{equation}\label{eq:eos}
  \sigma(\mathit{\Gamma}) = \sigma_0 + \mathcal{R}T\mathit{\Gamma}_\infty \ln\left(1 - \frac{\mathit{\Gamma}}{\mathit{\Gamma}_\infty}\right),
\end{equation}
where $\mathcal{R}$ is the ideal gas constant, $T$ is the temperature, $\sigma_0$ is the interfacial tension of the surfactant-free interface, and $\mathit{\Gamma}_\infty$ is the maximum packing concentration. Although this model is used in the numerical tests presented below, the proposed method is not restricted to this particular EOS.

Finally, the evolution of the surfactant concentration $\mathit{\Gamma}$ is governed by the advection-diffusion equation on the deformable, zero-thickness interface $\Upsilon$ \cite{stone_simple_1990}:
\begin{equation}\label{eq:sur-trans}
  \frac{\partial \mathit{\Gamma}}{\partial t} + \nabla_s \cdot (\mathit{\Gamma}\bm{u}) = \nabla_s \cdot (D_s \nabla_s \mathit{\Gamma}), \qquad \forall\bm{x} \in \Upsilon,
\end{equation}
where $D_s$ is the surface diffusivity of $\mathit{\Gamma}$ along the interface. The local surfactant mass is expressed as $M = \mathit{\Gamma} A$, and is obtained by integrating the interfacial surfactant concentration over the local control volume, i.e. the discretization cell in what follows.

In summary, the insoluble surfactant-laden flow is governed by Eqs.~(\ref{eq:ns1}) - (\ref{eq:sur-trans}) (supplemented by the linear dependence of $\rho$ and $\mu$ with respect to $c$). This set of equations constitutes a strongly coupled system for the fluid flow, interface evolution, and transport of the insoluble surfactant.

\subsection{Current numerical approaches}

Solving numerically the governing equations presented in \S~\ref{sec:governing-equations} introduces significant challenges, particularly when discretizing Eq.~(\ref{eq:sur-trans}). This equation holds on a zero-thickness interface evolving and deforming over time, a situation presenting difficulties to resolve accurately the advection and diffusion terms while maintaining the sharpness of the transport process. Additionally, numerical methods must ensure a strict conservation of the total surfactant mass, and accurately estimate the discretized interfacial mass $M$ and area $A$, as the local concentration is nothing but the ratio $\mathit{\Gamma} = M / A$ in the discretized cell. Achieving both mass conservation and accurate interface concentration, however, can represent conflicting goals when designing an overall computational strategy.

As discussed in \S\,\ref{sec:introduction}, existing numerical methods can be classified into two main categories: \textit{Eulerian}-based diffusive interface methods and \textit{Lagrangian}-based sharp methods. These approaches adopt distinct strategies to address the aforementioned challenges, as illustrated in Fig.~\ref{fig:schema}(\textit{b}) and (\textit{c}), respectively. In \textit{Eulerian}-based methods, directly discretizing Eq.~(\ref{eq:sur-trans}) on the zero-thickness interface proves challenging due to the dynamic and deformable nature of the latter. To circumvent this issue, diffusive interface methods introduce a finite-thickness region, $\widetilde{\Upsilon}$, around the interface (see Fig.~\ref{fig:schema}(\textit{b})). The surfactant concentration is transported within this finite-thickness region, allowing the use of standard 2D or 3D discretization schemes for advection-diffusion equations. This approach leads to a modified form of Eq.~(\ref{eq:sur-trans}), given as \cite{xu_level-set_2018}
\begin{equation}\label{eq:diffused-interface-method}
  \frac{\partial\mathit{\Gamma}}{\partial t} + \bm{u}\cdot\nabla\mathit{\Gamma} - \mathit{\Gamma}(\bm{n}\cdot\nabla)\bm{u}\cdot\bm{n} = D_s \left(\nabla^2\mathit{\Gamma} - \frac{\partial^2\mathit{\Gamma}}{\partial n^2} - \kappa \frac{\partial \mathit{\Gamma}}{\partial n}\right), \qquad \forall\bm{x}\in\widetilde{\Upsilon}\,,
\end{equation}
with $\partial/\partial n\equiv\bm{n}\cdot\nabla$. This equation can be discretized using standard finite difference methods. Nevertheless, the numerical robustness of this diffusive approach is obtained at the expense of the preservation of the interface sharpness, resulting in a lower numerical accuracy compared to that offered by sharp interface methods \cite{cleret_de_langavant_level-set_2017}.

Conversely, \textit{Lagrangian}-based sharp interface methods maintain the sharpness of the interface by tracking the surfactant concentration using Lagrangian markers (Fig.~\ref{fig:schema}(\textit{c})). More specifically, the surfactant concentration is stored at discrete markers along the interface, allowing Eq.~(\ref{eq:sur-trans}) to be reformulated in the Lagrangian framework in the form of a conservation equation for the surfactant mass, namely
\begin{equation}\label{eq:lagrangian}
  \frac{\mathrm{d} M}{\mathrm{d}t}  = \frac{\mathrm{d}(\mathit{\Gamma}A)}{\mathrm{d}t} =   A \nabla_s\cdot(D_s\nabla_s\mathit{\Gamma}),   \qquad \forall\bm{x}\in\Upsilon.
\end{equation}
Equation~\eqref{eq:lagrangian} avoids the complex discretization of the advection term in Eq.~(\ref{eq:sur-trans}), since $A$ can be obtained directly from the position of the Lagrangian markers. Besides, the diffusion term can be discretized using a second-order finite-difference of finite-volume scheme, guaranteing mass conservation. This method has proven effective in modeling surfactant-laden bubbly flows \cite{lu2017effect}. However, its application to 3D flows with substantial shape deformation and complex topological changes remains an open challenge.

Based on the above, it appears desirable to develop a numerical approach that preserves interface sharpness while accommodating topological changes such as those involved in bubble/drop coalescence and break-up. To the best of our knowledge, James and Lowengrub \cite{james_surfactant-conserving_2004} were the first, and likely the only ones, to propose a sharp interface method for solving insoluble surfactant transport along a deforming interface within an \textit{Eulerian}-based VOF framework. Instead of solving Eq.~(\ref{eq:sur-trans}), they solved Eq.~\eqref{eq:lagrangian} supplemented by the transport equation for the interfacial area $A$, namely
\begin{equation}\label{eq:james2}
  \frac{\partial A}{\partial t} + \bm{u}\cdot\nabla A  = -A(\bm{n}\cdot\nabla)\bm{u}\cdot\bm{n},   \qquad \forall\bm{x}\in\Upsilon.
\end{equation}
and obtained the surfactant concentration as $\mathit{\Gamma} = M/A$. We write symbolically Eqs. \eqref{eq:lagrangian}  and \eqref{eq:james2} in the form $\partial_t M = \mathcal{F}(M,\bm{u})$ and $\partial_t A = \mathcal{G}(A,\bm{u})$, respectively. The key issue with \eqref{eq:james2} is that it does not incorporate the information resulting from the VOF advection equation (\ref{eq:vof-adv}). As a result, the computed interface area $A$ is not consistent with the geometrically reconstructed interface area, leading to error accumulation over time. This discrepancy makes the results significantly deviate from physical reality in unsteady, long-duration simulations. Furthermore, extending the discretization schemes from \cite{james_surfactant-conserving_2004}, developed for a 2D axisymmetric system, to 3D coordinates is non-trivial.

Given the limitations of both diffusive and sharp interface methods, we propose in what follows a sharp and conservative numerical approach for simulating interfacial flows with insoluble surfactants within a geometrical VOF framework. This method establishes a self-consistent coupling between volume fraction and interface area advection routines, enabling higher-order interface reconstruction while ensuring compatibility with both 2D and 3D systems without extensive modifications. The remainder of this manuscript is organized as follows. In \S~\ref{sec:two-phase-flow-solver}, we describe the two-phase flow solver for Eqs.~(\ref{eq:ns1}) - (\ref{eq:body-force}), including the geometrical VOF approach and Marangoni force calculation. In \S~\ref{sec:sharp-scheme-surfactant-transport}, we detail the numerical techniques for discretizing Eq.~(\ref{eq:sur-trans}), with a focus on the treatment of both advection and diffusion terms.

\section{Two-phase flow solver}\label{sec:two-phase-flow-solver}

The two-phase flow solver is based on the open-source library \textit{Basilisk} \cite{popinet2024basilisk}, which we briefly review here. \S~\ref{sec:flow-solver} outlines the primary features of the numerical methods employed in \textit{Basilisk}, \S~\ref{sec:volume-fluid-vof} details the associated geometrical volume-of-fluid (VOF) method, and \S~\ref{sec:marang-force-calc} discusses the computation of the Marangoni force using the height-function (HF) method \cite{seric_direct_2018}. Furthermore, this section highlights improvements brought to the HF-based model in the present work. The solver also incorporates a quadtree/octree adaptive mesh refinement (AMR) technique, which significantly enhances computational efficiency by dynamically refining the grid near the interface, where high resolution is essential.

\subsection{Main feature of the flow solver}\label{sec:flow-solver}

To solve the Navier-Stokes equations (\ref{eq:ns1}) and (\ref{eq:ns2}), \textit{Basilisk} employs a second-order fractional step strategy \cite{chorin_numerical_1968}. This method allows for the coupling of the pressure field $p$ and velocity field $\bm{u}$ within a finite-volume framework. A staggered time discretization strategy is used, where pressure $p$ and volume fraction $c$ are computed at time steps $(n+1/2)$, while the velocity $\bm{u}$ is updated at time steps $(n + 1)$. The advection term $(\bm{u} \cdot \nabla)\bm{u}$ is discretized at time steps $(n+1/2)$ using the second-order upwind Bell-Colella-Glaz scheme \cite{bell_second-order_1989}. The viscous term $\nabla \cdot (\mu (\nabla \bm{u} + \nabla \bm{u}^T))$ is discretized semi-implicitly at time steps $(n+1/2)$ with a second-order central difference scheme, allowing the use of larger time steps. The interface mean curvature $\kappa$ and normal unit vector $\bm{n}$ involved in the normal component of the interfacial tension force, $\sigma \kappa  {\bm{n}} \delta_s$, are computed using a second-order accurate height-function (HF) method. A balanced-force model \cite{francois2006balanced} is implemented to minimize spurious currents near the free surface when discretizing the interfacial tension. These numerical technique have been extensively documented and validated in prior works \cite{popinet_accurate_2009, popinet_gerris_2003}, and further details are omitted here for brevity. 

\subsection{Geometrical VOF method}\label{sec:volume-fluid-vof}

\textit{Basilisk} employs a geometrical volume-of-fluid (VOF) method to compute the volume fraction $c$, governed by Eq.~(\ref{eq:vof-adv}). The interface $\Upsilon$ is subsequently updated and reconstructed based on the computed values of $c$. This method comprises two key features: (1) in solving Eq.~(\ref{eq:vof-adv}), the computation of the volume fraction flux across cell faces relies on the geometrical information concerning the reconstructed interface; and (2) a piecewise linear interface construction (PLIC) scheme \cite{gueyffier1999volume} is employed to reconstruct the interface once the volume fractions are updated.

To numerically solve the transport equation for $c$, a direction-splitting scheme \cite{popinet_accurate_2009} is adopted. Since the velocity field $\bm{u}$ does not strictly satisfy the divergence-free condition in individual directions, Eq.~(\ref{eq:vof-adv}) is rewritten as
\begin{equation}\label{eq:vof-adv2}
  \frac{\partial c}{\partial t} + \nabla\cdot(\bm{u}c) - c\nabla\cdot\bm{u} = 0.
\end{equation}

\begin{figure}
  \centering
  \includegraphics[width=.9\textwidth]{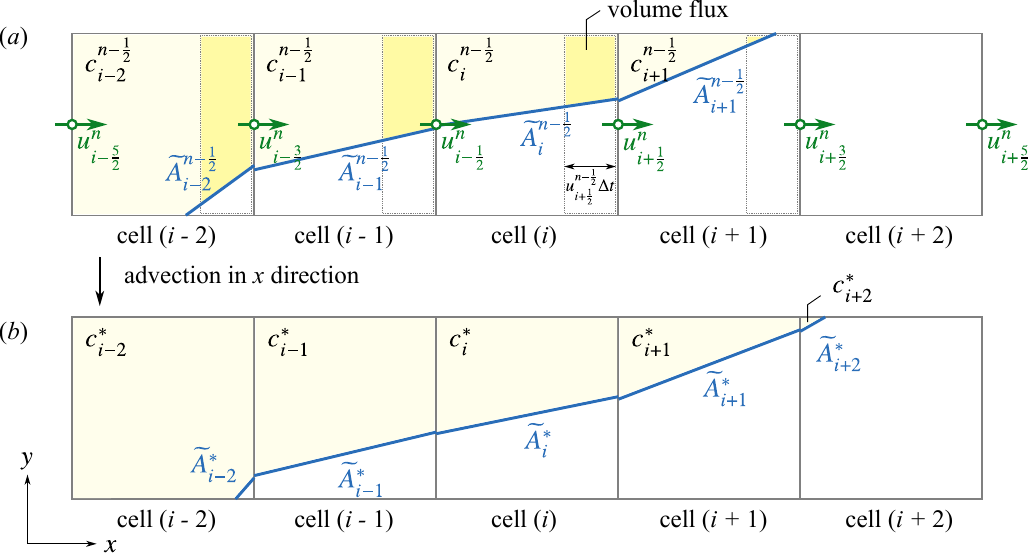}
  \caption{Illustration of a volume-of-fluid advection step in the $x$-direction for a 2D case at time layer $(n-1/2)$, corresponding to Eq.~(\ref{eq:vof-adv-discre-1}). Rows (\textit{a}) and (\textit{b}) show the interfaces before and after the advection step, respectively. The light yellow and white regions represent phases 1 and 2, respectively. The dark yellow region illustrates the advection flux moving into the cell on the right. The volume of phase 1 within a cell defines the local volume fraction $c$. The solid blue lines depict the reconstructed interfaces, with their area denoted by $\widetilde{A}$.}
  \label{fig:advection}
\end{figure}

For clarity, we detail the discretization process in the 2D case. Since $c$ is computed at staggered time steps, its update in cell $(i, j)$ from $(n - 1/2)$ to $(n+1/2)$ is formulated as
\begin{equation}\label{eq:vof-adv-discre-1}
  \mathit{x-}\mathrm{direction}: \qquad \frac{c^*_{i,j} - c^{n-\frac12}_{i,j}}{\Delta t} + \frac{(uc)^{n-\frac12}_{i+\frac12,j}-(uc)^{n-\frac12}_{i-\frac12,j}}{\Delta}-c_c\frac{u^{n-\frac12}_{i+\frac12,j}-u^{n-\frac12}_{i-\frac12,j}}{\Delta} = 0,
\end{equation}
\begin{equation}\label{eq:vof-adv-discre-2}
  \mathit{y-}\mathrm{direction}: \qquad   \frac{c^{n+\frac12}_{i,j} - c^*_{i,j}}{\Delta t} + \frac{(vc)^*_{i,j+\frac12}-(vc)^*_{i,j-\frac12}}{\Delta}-c_c\frac{v^{n-\frac12}_{i,j+\frac12}-v^{n-\frac12}_{i,j-\frac12}}{\Delta} = 0
\end{equation}
for the $x$ and $y$ directions, respectively. Here, $u$ and $v$ represent the velocity components of $\bm{u}$ in the $x$ and $y$ directions, respectively. The subscript $(i, j)$ indicates that the variable is located at the cell center, while $(i \pm 1/2, j)$ and $(i, j \pm 1/2)$ refer to face-centered variables (we sometimes use the subscript $f$ to denote them generically). The superscript $*$ denotes intermediate values of $c$. The same notation will later be applied to other scalars discussed below. Following Weymouth and Yue \cite{weymouth_conservative_2010}, the volume fraction $c_c$ involved in the dilatation term of \eqref{eq:vof-adv-discre-1} and \eqref{eq:vof-adv-discre-2} is set to 1 (or 0) if $c^{n-1/2}_{i,j} > 0.5$ ($\leq 0.5$) to avoid overfilling ($c > 1$) or over-emptying ($c < 0$) the corresponding cell. Figure \ref{fig:advection}(\textit{a}) illustrates the geometrical approach used to compute the flux $(uc)^{n-1/2}_f \Delta t$ across cell faces (the $j$ indices are omitted for simplicity). The dark yellow regions represent the advected volume fraction of phase 1 ($c > 0$) moving to the right neighbor in the $x$-direction. Using the known displacement $u_f \Delta t$ and the geometrical information of the interface $\Upsilon$, the volume of these regions is calculated, and the flux $(uc)^{n-1/2}_f \Delta t$ is substituted into Eq.~(\ref{eq:vof-adv-discre-1}) to update $c^*_{i,j}$. A PLIC scheme is then applied to reconstruct $\Upsilon^*$ as line segments (2D) or facets (3D), as depicted in Fig.~\ref{fig:advection}(\textit{b}). The procedure is immediately repeated for the $y$-direction, solving Eq.~(\ref{eq:vof-adv-discre-2}) to compute $c^{n+1/2}$ and reconstruct $\Upsilon^{n+1/2}$. Note for future purpose that the interface area $\widetilde{A}$ reconstructed with the PLIC technique differs from the interface area $A$ used in the concentration calculation. This method extends naturally to 3D cases, though the computation of geometrical fluxes is more complex; the reader is referred to \cite{scardovelli_analytical_2000} for further details.

The VOF method ensures volume conservation of both phases while maintaining a sharp interface through the geometrical PLIC scheme. These strengths make \textit{Basilisk} accurate and versatile for general two-phase flow simulations. However, a significant limitation arises when applying this geometrical method to surfactant-laden flows. The PLIC scheme reconstructs the interface and obtains the discretized area $\widetilde{A}$ based on the conservation of the volume fraction, making $\widetilde{A}$ differ from the real interface area $A$ within the discretization cell. While this difference has negligible impact on the flow dynamics, it introduces significant errors when computing the surfactant concentration $\mathit{\Gamma}$, which depends on both the surfactant mass $M$ and the interfacial area $A$. To address this issue, the present work introduces a novel numerical method for accurately computing $\mathit{\Gamma}$ within the framework of the geometrical VOF scheme, as described in \S~\ref{sec:sharp-scheme-surfactant-transport}.

\subsection{Marangoni force calculation}\label{sec:marang-force-calc}

\begin{figure}
  \centering
  \includegraphics[width=.9\textwidth]{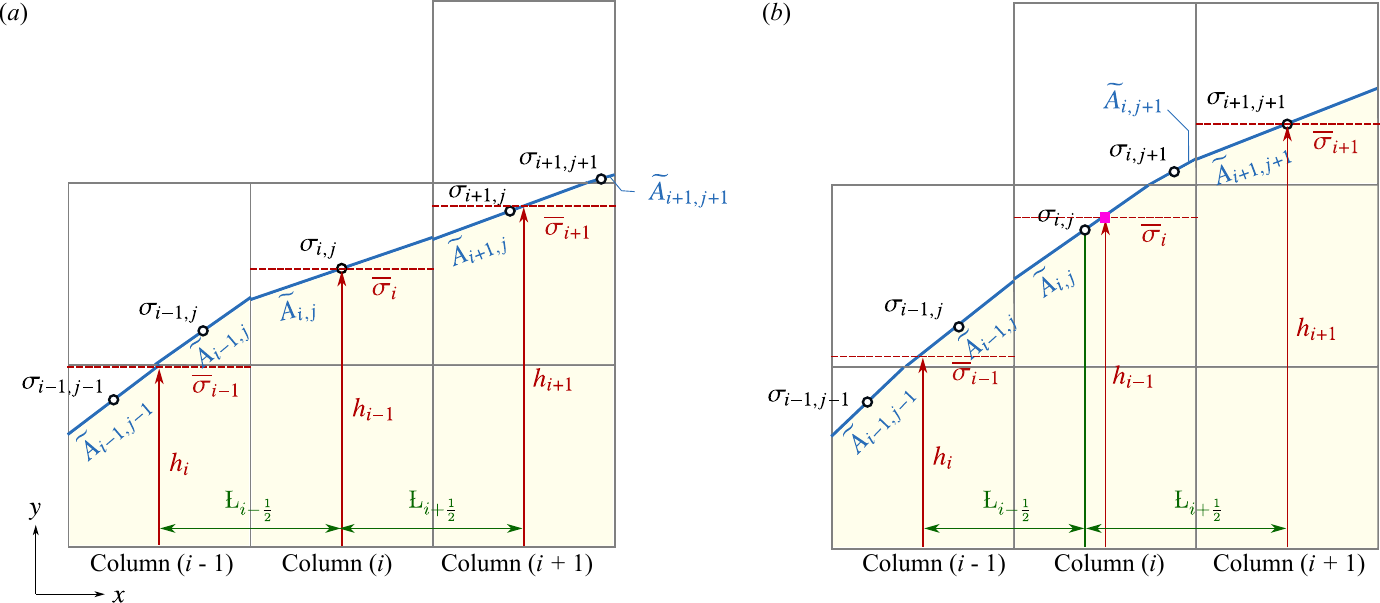}
  \caption{Schematic illustration of the numerical schemes adopted for the Marangoni force calculation. The variable interfacial tension $\sigma$ is defined at the middle of the reconstructed interface segment. Through the height-function method (HFM), the height function $h$ and the effective interfacial tension $\overline \sigma$ are calculated for each column. (\textit{a}) represents the case where the middle of the reconstructed interface segment in cell $(i,j)$ lies on the cell's vertical central line, while (\textit{b}) represents the case where it does not. As discussed in the text, different schemes are adopted for the Marangoni force calculation in cell $(i,j)$ in cases (\textit{a}) and (\textit{b}). }
  \label{fig:marangoni-force-calc}
\end{figure}

The numerical model for the normal interfacial tension force, $\sigma\kappa \bm{n} \delta_s$, in \textit{Basilisk} incorporates a balanced-force model \cite{francois2006balanced} based on the height-function (HF) technique \cite{cummins2005estimating}. Numerical details of this implementation are thoroughly discussed in \cite{popinet_accurate_2009}. Consequently, we focus here on the numerical method used for computing the tangential component, $\nabla_s\sigma \delta_s$, induced by interfacial tension gradients, which corresponds to the Marangoni force arising from surfactant-induced inhomogeneities in $\sigma$. While the original algorithm was proposed by Seric \textit{et al.} \cite{seric_direct_2018} for thermocapillary flows, the present work introduces several improvements to extend the method to surfactant-laden interfaces.

Figure~\ref{fig:marangoni-force-calc}(\textit{a}) illustrates the way $\nabla_s \sigma$ is estimated within an interfacial cell $(i,j)$. The computation relies on HF values from the local column $i$ and its two neighboring columns, $(i \pm 1)$, denoted as $h_{i-1}$, $h_{i}$, and $h_{i+1}$. These HF values are pre-computed from the volume fractions in the respective columns \cite{popinet_accurate_2009, cummins2005estimating}. For example, the HF value in column $i$ is obtained by summing the volume fractions along the $y$-direction, leading to $h_i = \sum_j c_{i, j} \Delta$, where the summation extends over a maximum of nine cells within the column, typically from $(i, j-4)$ to $(i, j+4)$ in Fig.~\ref{fig:marangoni-force-calc}(\textit{a}), unless cells further away are empty. Next, effective interfacial tensions, $\overline\sigma_{i-1}$, $\overline{\sigma}_{i}$, and $\overline{\sigma}_{i+1}$, are computed for these columns using an area-weighted averaging scheme. For instance, $\overline{\sigma}_{i-1}$ is interpolated from the interfacial tension values in the two interfacial cells, $(i-1,j-1)$ and $(i-1,j)$, of column $(i-1)$, according to
\begin{equation}
  \overline{\sigma}_{i-1} = \frac{\widetilde{A}_{i-1,j-1}\sigma_{i-1,j-1}+ \widetilde A_{i-1,j}\sigma_{i-1,j}}{\widetilde A_{i-1,j-1}+\widetilde A_{i-1,j}}.
\end{equation}

The same procedure is applied to column $(i+1)$. For column $i$, where only one interfacial cell exists, we set directly $\overline{\sigma}_{i} = \sigma_{i,j}$. This area-weighted averaging scheme is preferred to the fraction-weighted approach proposed by Seric \textit{et al.} \cite{seric_direct_2018}, as it has proven more robust in practice. Using the HF values and the above effective interfacial tensions, $\nabla_s \sigma$ is estimated in the interfacial cell $(i,j)$. For the case depicted in Fig.~\ref{fig:marangoni-force-calc}(\textit{a}), a central difference scheme is employed, namely
\begin{equation}\label{eq:surface-gradient-aligned}
  \left(\nabla_s\sigma\right)_{i,j} = \frac{\overline{\sigma}_{i+1} - \overline{\sigma}_{i-1}}{2\Delta\sqrt{1+(\partial_xh|_{i,j})^2}}\bm{\tau}_{i,j},
\end{equation}
where $\partial_x h |_{i,j}$, the $x$-component derivative of the HF at the interface centroid of cell $(i,j)$, is discretized as $\partial_x h |_{i,j} \approx (h_{i+1} - h_{i-1}) / 2\Delta$. Besides, the unit tangent vector $\bm{\tau}_{i,j}$ is calculated as: $\bm{\tau}_{i,j} = (1, \partial_x h |_{i,j}) / \sqrt{1+(\partial_x h |_{i,j})^2}$. This central difference scheme achieves second-order accuracy only if the centroid of the interface segment in column $i$ lies on the vertical centerline, such that $\L_{i-1/2} =\L_{i+1/2} = \Delta$, as shown in Fig.~\ref{fig:marangoni-force-calc}(\textit{a}). However, when this centroid deviates from the vertical centerline, as illustrated in Fig.~\ref{fig:marangoni-force-calc}(\textit{b}) where two interface segments are involved in the central column $i$, Eq.~(\ref{eq:surface-gradient-aligned}) becomes inaccurate. In such cases, the computed $\left(\nabla_s \sigma\right)_{i,j}$ estimated from Eq.~(\ref{eq:surface-gradient-aligned}) represents a second-order approximation at the vertical centerline (magenta square marker in Fig.~\ref{fig:marangoni-force-calc}(\textit{b})), but not at the actual centroid of the interface segment (circular marker). To correct this shift, we generalize \eqref{eq:surface-gradient-aligned} to cases where $\L_{i-1/2} \neq \L_{i+1/2}$ in the form
\begin{equation}\label{eq:surface-gradient-not-aligned}
  \left(\nabla_s\sigma\right)_{i,j} \approx \frac{(\overline{\sigma}_{i+1}-\sigma_{i,j})\L_{i-\frac12}^2 - (\overline{\sigma}_{i-1}-\sigma_{i,j})\L_{i+\frac12}^2}{2\Delta\L_{i-\frac12}\L_{i+\frac12}\sqrt{1+(\partial_xh|_{i,j})^2}}\bm{\tau}_{i,j}.
\end{equation}

With this formulation, the tangential Marangoni force is computed using the vertical HF values $h(y)$, and a similar procedure can be applied to rows using horizontal HF values $h(x)$, depending on the slope of the reconstructed interfaces.

\subsection{Time step constraints}

The time step is determined by the constraints arising from the advective and capillary terms involved in the governing equations. First, in the VOF advection equation, the time step is restricted by the standard Courant-Friedrichs-Lewy (CFL) condition imposing that $\Delta t \leq c_{\mathrm{CFL}}~\Delta_{\mathrm{min}}/\mathrm{max}(\|\bm{u}\|)$, where $c_{\mathrm{CFL}}$ is a user-defined parameter ($c_{\mathrm{CFL}} < 0.5$) and $\Delta_{\mathrm{min}}$ is the finest cell size produced by the AMR procedure. Second, the capillary term in the momentum equation is advanced explicitly in time. Stability then requires that the time step be smaller than the period of the shortest capillary wave that can be captured by the grid \cite{popinet_accurate_2009}, \textit{i.e.} $\Delta t \leq \sqrt{\rho_m \Delta_{\mathrm{min}}^3/ \pi \sigma_{\mathrm{max}}}$, with $\rho_m = (\rho_1+\rho_2)/2$ and $\sigma_{\mathrm{max}}$ being the maximum interfacial tension over the entire interface. Finally, the time step to be used in the simulation is selected as
\begin{equation}\label{eq:time-step}
 \Delta t = \mathrm{min}\left[\frac{c_{\mathrm{CFL}}~\Delta_{\mathrm{min}}}{\mathrm{max}(\|\bm{u}\|)}, \sqrt{\frac{\rho_m \Delta_{\mathrm{min}}^3}{ \pi \sigma_{\mathrm{max}}}}\right]\,.
\end{equation}

\section{Sharp numerical method for the transport of insoluble surfactants}\label{sec:sharp-scheme-surfactant-transport}

In this section, we detail the sharp interface method we developed for solving the transport equation of insoluble surfactants along reconstructed interfaces within the geometric VOF framework. The main numerical challenges encountered in the development of such a method in the framework of an \textit{Eulerian} approach stem from the discretization of the advection and diffusion terms in Eq.~(\ref{eq:sur-trans}). These challenges are particularly demanding due to the intricate coupling between the surfactant transport and the evolution of the zero-thickness interface. Maintaining physical consistency in this coupling is essential for achieving accurate simulations.

\begin{figure}
  \centering
  \includegraphics[width=.6\textwidth]{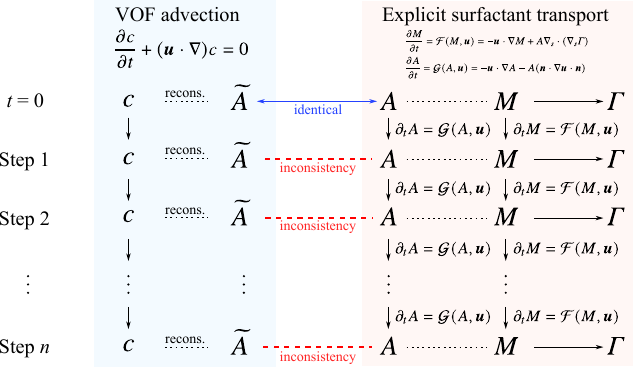}
  \caption{Illustration of the source of inconsistency between the area of the reconstructed interface, $\widetilde{A}$, and the transported area, $A$, in explicit approaches such as  \cite{james_surfactant-conserving_2004}. Although the transported interface area $A$ is identical to the reconstructed area $\widetilde{A}$ at $t=0$, the two follow different evolutions: $A$ is governed by $\partial_tA = \mathcal{G}(A, \bm{u})$, while $\widetilde{A}$ is dictated by the VOF advection scheme.}
  \label{fig:surfactant-adv-explicit}
\end{figure}

Being formulated in the framework of diffuse interface methods \cite{xu_eulerian_2003}, Eq.~(\ref{eq:diffused-interface-method}) is not directly applicable here because the high-order terms it contains can hardly be discretized along a zero-thickness interface. We do not adopt the approach of \cite{james_surfactant-conserving_2004} either, which splits the concentration equation into two separate equations governing $M$ and $A$, respectively. This method fails to establish a direct connection between the advection equation for the volume fraction and the advection-stretching equation for the interface area, leading to inconsistent estimates of the latter. This inconsistency stems from the overdetermined nature of the interface area, which is treated both as an explicitly transported quantity governed by $\partial_t A = \mathcal{G}(A,\bm{u})$, and an implicitly reconstructed area $\widetilde{A}$ stemming from the VOF advection process. This inconsistency is illustrated in Fig.~\ref{fig:surfactant-adv-explicit}, based on the approach of \cite{james_surfactant-conserving_2004}.

\begin{figure}
  \centering
  \includegraphics[width=\textwidth]{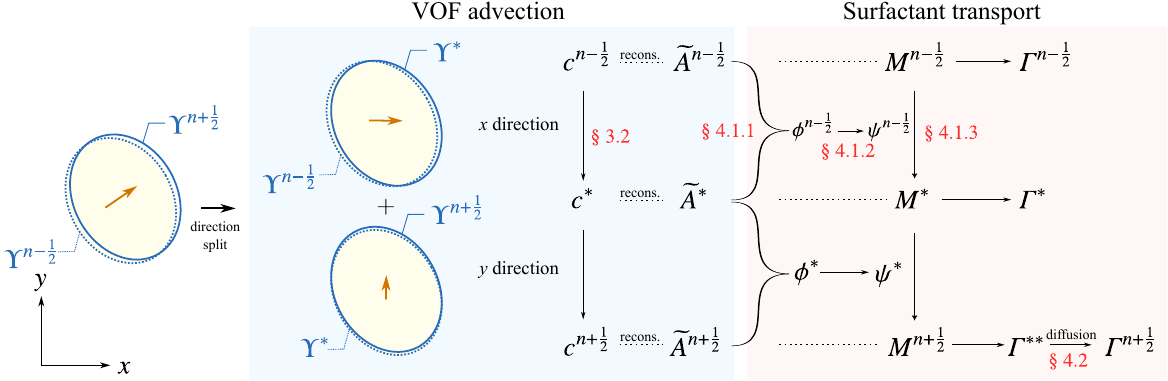}
    \vspace{-3mm}
  \caption{Global strategy used to solve the surfactant transport equation along the VOF reconstructed interface within the time step $(n-1/2,n+1/2)$. The illustration is for the 2D case; its extension to the 3D case is straightforward. Locations in \S~\ref{sec:sharp-scheme-surfactant-transport} where the the detailed description of each step of the procedure is given are highlighted in red.}
  \label{fig:global-procedures}
\end{figure}

To remove this inconsistency, we propose an original strategy that differs fundamentally from existing Eulerian approaches. Specifically, instead of solving the explicit equations \eqref{eq:lagrangian}-\eqref{eq:james2} governing $M$ and $A$, we still consider the concentration equation \eqref{eq:sur-trans} but also leverage on the VOF advection step to extract the implicit information required to accurately estimate the interface area. This approach leads to a more consistent representation of surfactant transport during the interface reconstruction within a VOF advection step, as illustrated in Fig.~\ref{fig:advection}. Within a single time step, Eq.~(\ref{eq:sur-trans}) is split into an advection step and a diffusion step, so that it is reformulated as
\begin{equation}\label{eq:aa1}
  \frac{\mathit{\Gamma}^{**} - {\mathit{\Gamma}^{n - \frac12}}}{\Delta t}  + \nabla_s\cdot({\mathit{\Gamma}}\bm{u})= 0, \qquad \forall\bm{x}\in{\Upsilon},
\end{equation}
\begin{equation}\label{eq:aa2}
  \frac{\mathit{\Gamma}^{n + \frac12} - \mathit{\Gamma}^{**}}{\Delta t}  = \nabla_s\cdot(D_s\nabla_s\mathit{\Gamma}),  \qquad \forall\bm{x}\in{\Upsilon},
\end{equation}
where $\mathit{\Gamma}^{**}$ is the intermediate surfactant concentration after the advection step. We will show later that $\mathit{\Gamma}^{**}$ is still estimated as $\mathit{\Gamma}^{**} = M^{**}/A^{n+1/2}$ at the end of the advection step, with the surfactant mass, $M^{**}$, and interface area, $A^{n+1/2}$, being both computed in accordance with the VOF advection procedure. 
The diffusion step then follows, completing the update of the concentration field. Figure \ref{fig:global-procedures} illustrates the proposed numerical strategy used to transport a surfactant in 2D along the reconstructed VOF interface within a single time step, $(n - 1/2) -(n + 1/2)$. We detail the advection and diffusion steps in \S\S~\ref{sec:cons-advect-meth} and \ref{sec:geom-volume-fram}, respectively.

\subsection{Sharp scheme for the advection step}\label{sec:cons-advect-meth}

In the advection step, Eq.~(\ref{eq:aa1}) is discretized successively in the $x$ and $y$ directions, aligning with the VOF direction-splitting scheme. Taking the time-space integral of (\ref{eq:aa1}) in the interfacial cell $(i, j)$ yields
\begin{equation}\label{eq:aa3}
  \mathit{x-}\mathrm{direction:} \qquad \widehat M^*_{i, j} - M^{n - \frac12}_{i, j}  + \psi^{n-\frac{1}{2}}_{i + \frac12, j} - \psi^{n-\frac{1}{2}}_{i - \frac12, j} = 0, \qquad \forall\bm{x}\in{\Upsilon},
\end{equation}
\begin{equation}\label{eq:aa4}
  \mathit{y-}\mathrm{direction:} \qquad   \widehat M^{**}_{i, j} -  \widehat M^*_{i, j}  + \psi^{*}_{i, j + \frac12} - \psi^{*}_{i, j - \frac12} = 0, \qquad \forall\bm{x}\in{\Upsilon},
\end{equation}
where $\psi$ represents the mass flux of the surfactant that crosses the corresponding cell face over time $\Delta t$. The notation $\,\widehat\cdot\,$ indicates that the calculated mass is provisional, and additional procedures are required to finalize the transitions from $\widehat{M}^*$ to $M^*$ and from $\widehat{M}^{**}$ to $M^{**}$. These procedures will be detailed in \S~\ref{sec:update-surf-conc}. Accurately solving the direction-splitting advection equations (\ref{eq:aa3}) and (\ref{eq:aa4}) requires consistency with the geometrical advection scheme used in the VOF equation. Additionally, the computation of the interface area $A$ must incorporate the stretching/compression effects implicitly embedded in the VOF solution. Starting along the $x$ direction, the evaluation of the mass flux $\psi^{n-1/2}$ must align with the motion of the deforming interface, which has an area $A^*$ determined from the intermediate volume fraction $c^*$. To achieve this, the area flux $\phi^{n-1/2}$ at the common boundary between two successive interfacial cell must be computed first. The area flux represents the length (in 2D) or area (in 3D) of the interface crossing the cell face as the interface area transitions from $A^{n-1/2}$ to $A^*$. The numerical scheme employed for determining $\phi^{n-1/2}$ is detailed in \S~\ref{sec:area-flux-estimation}. Once $\phi^{n-1/2}$ is computed, the corresponding mass flux of surfactant, $\psi^{n-1/2}$, is calculated following the process explained in \S~\ref{sec:mass-flux-estimation}. Subsequently, using $\psi^{n-1/2}$, the surfactant mass is updated from $M^{n-1/2}$ to $ M^*$  thanks to the numerical scheme outlined in \S~\ref{sec:update-surf-conc}.

This procedure is repeated along the $y$ direction, resulting in the updated interface area $A^{n+1/2}$ and surfactant mass $M^{**}$. It will be established in \S~\ref{sec:area-flux-estimation} that the correct $A$ is neither the area $\widetilde A$ obtained from PLIC reconstruction, nor that obtained by merely solving an advection-stretching equation \cite{james_surfactant-conserving_2004}. Instead, $A$ is implicitly computed when solving the VOF advection equation, in order to maintain numerical consistency. Surfactant concentration can then be advected, as illustrated in Fig.~\ref{fig:global-procedures}, yielding the intermediate surfactant concentration $\mathit{\Gamma}^{**}$.

\subsubsection{Area flux calculation }\label{sec:area-flux-estimation}

\begin{figure}
  \centering
  \includegraphics[width=\textwidth]{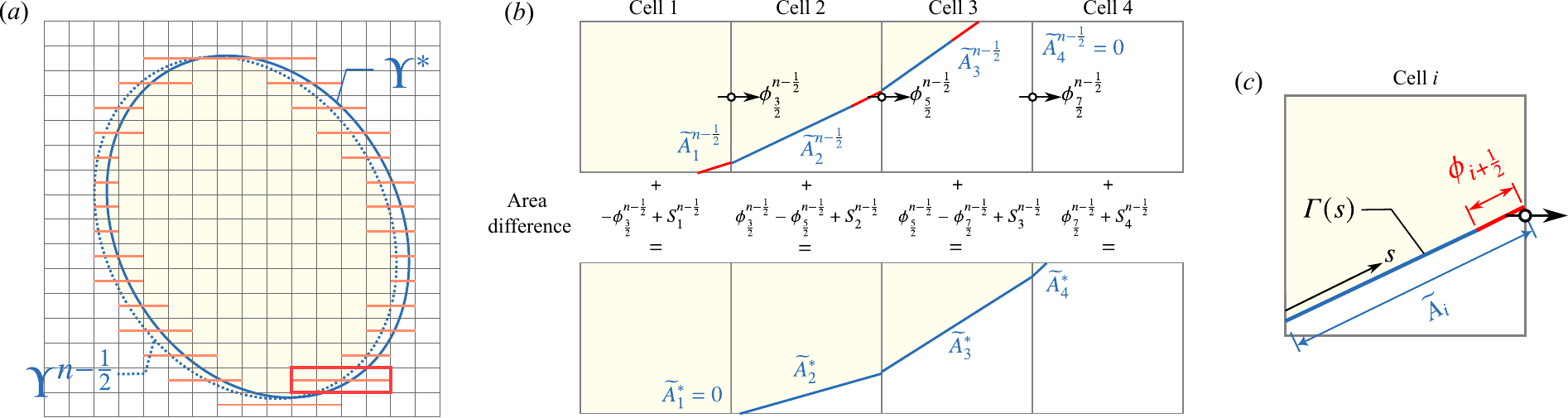}
    \vspace{-2mm}
  \caption{(\textit{a}) Definition of the cell panel (cells crossed by the orange line segments) in the $x$ direction. (\textit{b}) Variation of the reconstructed area $\widetilde A$ before and after the VOF advection in the $x$ direction in the `cell panel' highlighted near the bottom-right corner of (\textit{a}).
    (\textit{c}) Calculation of the mass flux (mass contained in the red portion) at the cell face $5/2$ in (\textit{b}) as an example.}
  \label{fig:area-flux-estimation}
\end{figure}

The area flux $\phi^{n-1/2}$, associated with the difference in interface area $A^* - A^{n-1/2}$ after and before updating the volume fraction in the $x$ direction, is implicitly contained in the solution of Eq.~(\ref{eq:vof-adv-discre-1}), the VOF equation along the same direction. The numerical techniques employed to compute $\phi^{n-1/2}$ are illustrated in Fig.~\ref{fig:area-flux-estimation}. Panel (\textit{a}) depicts the motion and deformation of the interface in the $x$ direction, with the dotted and solid lines representing interfaces $\Upsilon^{n-1/2}$ and $\Upsilon^{*}$, respectively. The orange segments indicate the cells forming the panel used to estimate $\phi^{n-1/2}$ in each horizontal row. During this sub-step, the interface does not cross the far-left and far-right cell faces. Fig.~\ref{fig:area-flux-estimation}(\textit{b}) provides a zoomed-in view taken near the bottom-right corner of panel (\textit{a}). The top row in Fig.~\ref{fig:area-flux-estimation}(\textit{b}) corresponds to the time step $(n-1/2)$, while the bottom row represents the intermediate step $*$. Comparing both rows, it may be noticed that the left face of cell 1 remains fully occupied by fluid 1 during this stage, while the right face of cell 4 remains completely empty, implying that the area fluxes at these faces, $\phi^{n-1/2}_{1/2}$ and $\phi^{n-1/2}_{9/2}$, are both zero. For clarity, the subscript $j$ is omitted in the following discussion, as we focus solely on discretization in the $x$ direction. Hence, cell $i$ actually refers to cell $(i, j)$.

The difference in interface area contained within cell $i$ before and after the advection step, computed from the geometric information resulting from the PLIC reconstruction can be expressed as
\begin{equation}\label{eq:area-diff}
  \widetilde A^*_i - \widetilde A_i^{n-\frac12} = \phi^{n-\frac12}_{i-\frac12} - \phi^{n-\frac12}_{i+\frac12} + \Delta t S^{n-\frac12}_i, \qquad \qquad i\in1, 2, 3, 4,
\end{equation}
where $S_i^{n-1/2}$ represents the source term accounting for the area change due to the compression/stretching effects involved in the unidirectional advection of the VOF equation, and $\phi^{n-1/2}_{i-1/2}$ and $\phi^{n-1/2}_{i+1/2}$ represent the area flux crossing cell faces $i-1/2$ and $i+1/2$, respectively. By convention, $\phi$ is considered positive (negative) when the area is transported from the left (right) cell to the right (left) cell. The exact value of $S_i^{n-1/2}$ is not directly known and we approximate it as
\begin{equation}\label{eq:strech-compress}
  S^{n-\frac12}_i = \widetilde A_i^{n-\frac12}(\nabla\cdot\bm{u}-\bm{n}\cdot\nabla\bm{u}\cdot\bm{n})_i^n + \varepsilon_{i}^{n-\frac12}, \qquad \qquad i\in1, 2, 3, 4,
\end{equation}
In \eqref{eq:strech-compress}, $\widetilde A_i^{n-1/2}(\nabla\cdot\bm{u}-\bm{n}\cdot\nabla\bm{u}\cdot\bm{n})_i^n \equiv \widetilde A_i^{n-1/2}[(\mathbb{I}-\bm{n}\bm{n}^T):\nabla\bm{u}]_i^n$ is the theoretical expression for the compression/stretching effect, and $\varepsilon_{i}^{n-1/2}$ is the unavoidable error term arising from the numerical estimation of the various terms. Equation (\ref{eq:area-diff}) thus becomes
\begin{equation}\label{eq:area-diff-2}
  \widetilde A^*_i - \widetilde A_i^{n-\frac12} = \phi^{n-\frac12}_{i-\frac12} - \phi^{n-\frac12}_{i+\frac12} + \Delta t[\widetilde A_i^{n-\frac12}(\nabla\cdot\bm{u}-\bm{n}\cdot\nabla\bm{u}\cdot\bm{n})_i^n + \varepsilon_{i}^{n-\frac12}]\,, \qquad \qquad i\in1, 2, 3, 4.
\end{equation}
Equation~(\ref{eq:area-diff-2}) alone is insufficient for estimating the area flux, $\phi^{n - 1/2}$. Indeed, although $\widetilde A_i^{n-1/2}$ and $\widetilde A^*$ are determined from the geometric reconstruction of the interface, three unknowns remain: $\phi^{n-1/2}_{i-1/2}$, $\phi^{n-1/2}_{i+1/2}$, and $\varepsilon_{i}^{n-1/2}$. To overcome this issue, we sum \eqref{eq:area-diff-2} over the entire panel, which yields
\begin{equation}\label{eq:12}
  \sum_{\ell = 1}^4 \left(\widetilde A^*_\ell-\widetilde A_\ell^{n-\frac12}\right) = \underbrace{\sum_{\ell = 1}^4\left(\phi^{n-\frac12}_{\ell-\frac12} - \phi^{n-\frac12}_{\ell+\frac12}\right)}_{=0} +\Delta t\sum_{\ell = 1}^4 \widetilde A_\ell^{n-\frac12}(\nabla\cdot\bm{u}-\bm{n}\cdot\nabla\bm{u}\cdot\bm{n})_\ell^n + \Delta t\sum_{\ell = 1}^4\varepsilon_{\ell}^{n-\frac12},
\end{equation}
in which the net area flux crossing the extreme left and right faces of the cell panel vanishes by construction. This property allows us to compute the sum of the unknown error term as
\begin{equation}\label{eq:13}
  \sum_{\ell = 1}^4\varepsilon_{\ell}^{n-\frac12}   =  (\Delta t)^{-1} \sum_{\ell = 1}^4 \left(\widetilde A^*_\ell-\widetilde A_\ell^{n-\frac12}\right) - \sum_{\ell = 1}^4 \widetilde A_\ell^{n-\frac12}(\nabla\cdot\bm{u}-\bm{n}\cdot\nabla\bm{u}\cdot\bm{n})_\ell^n\,.
\end{equation}

Equation (\ref{eq:13}) represents, for the given cell panel, the difference between the reconstructed interface area $\widetilde A$ and the interface area $A$ predicted by the VOF transport, in which stretching/compression effects are accounted for. To distribute this difference among the interfacial cells, we employ an area-weighted scheme. Thus, the source term $\widetilde{\varepsilon}_i^{n-\frac12}$ accounting for this difference in cell $i$ is approximated as
\begin{equation}\label{eq:14}
  \widetilde{\varepsilon}_i^{n-\frac12} = \frac{ \widetilde A^{n-\frac12}_i}{\sum_{\ell = 1}^4 \widetilde A^{n-\frac12}_\ell} \sum_{\ell = 1}^4\varepsilon_{\ell}^{n-\frac12}, \qquad \qquad i = 1, 2, 3, 4.
\end{equation}

Next, we insert \eqref{eq:14} in (\ref{eq:area-diff-2}) for each cell. This yields a matrix equation for $\phi^{n - 1/2}$ in the form
\begin{equation}\label{eq:15}
  \begin{pmatrix}
    1  & -1     &       &        &     &   \\
       & 1 & -1 &  &        &  \\
       &  & 1 & -1 & &    \\
       &  &  & \ddots & \ddots &  \\
       &        & & & 1 & -1 \\
  \end{pmatrix}
  \begin{pmatrix}
    0\\ \phi^{n-\frac12}_{\frac32}\\ \phi^{n-\frac12}_{\frac52}\\ \vdots \\ \phi^{n-\frac12}_{N-\frac12}\\0
  \end{pmatrix}
  \approx
  \begin{pmatrix}
    \widetilde A^*_1 - \widetilde A_1^{n-\frac12}[1+\Delta t (\nabla\cdot\bm{u}-\bm{n}\cdot\nabla\bm{u}\cdot\bm{n})_1^n] - \Delta t\widetilde{\varepsilon}_{1}^{n-\frac12}  \\
    \widetilde A^*_2 - \widetilde A_2^{n-\frac12}[1+\Delta t(\nabla\cdot\bm{u}-\bm{n}\cdot\nabla\bm{u}\cdot\bm{n})_2^n] - \Delta t\widetilde{\varepsilon}_{2}^{n-\frac12} \\
    \widetilde A^*_3 - \widetilde A_3^{n-\frac12}[1+\Delta t(\nabla\cdot\bm{u}-\bm{n}\cdot\nabla\bm{u}\cdot\bm{n})_3^n] - \Delta t\widetilde{\varepsilon}_{3}^{n-\frac12} \\
    \vdots \\
    \widetilde A^*_N- \widetilde A_N^{n-\frac12}[1+\Delta t(\nabla\cdot\bm{u}-\bm{n}\cdot\nabla\bm{u}\cdot\bm{n})_N^n] - \Delta t\widetilde{\varepsilon}_{N}^{n-\frac12}
  \end{pmatrix}\,,
\end{equation}
with $N=4$ in the case of the cell panel depicted in Fig.~\ref{fig:area-flux-estimation}. This sparse matrix system can be efficiently solved, completing the computation of the area flux $\phi^{n - 1/2}_{i + 1/2}$ across the faces of interfacial cells. The same approach can be applied to estimate the area flux on cell faces oriented in the $y$ direction.

Equation (\ref{eq:15}) establishes a relation between the reconstructed interface area $\widetilde{A}$ and the transport-predicted interface area through the area fluxes $\phi^{n-\frac12}$. Once these fluxes are known, a correction can be applied to $\widetilde{A}$ to properly estimate the interface area involved in the transport of the surfactant, ensuring consistency with the VOF advection scheme.

\subsubsection{Mass flux calculation}\label{sec:mass-flux-estimation}

Once the area flux $\phi^{n - 1/2}$ is determined at the cell faces, the corresponding mass flux of surfactant, $\psi^{n - 1/2}$, can be computed by evaluating the product of the local surfactant concentration $\mathit{\Gamma}^{n - 1/2}$ and area flux $\phi^{n - 1/2}$ in the same cell. At the cell face $i+1/2$, assuming a positive area flux $\phi^{n - 1/2}_{i+1/2}$ from cell $i$ to $i + 1$ as illustrated in Fig.~\ref{fig:area-flux-estimation}(\textit{c}), the mass flux $\psi^{n - 1/2}_{i+1/2}$ can straightforwardly be estimated as $\psi^{n - 1/2}_{i+1/2} = \phi^{n - 1/2}_{i+1/2}\mathit{\Gamma}^{n - 1/2}_i$, where $\mathit{\Gamma}^{n - 1/2}_i$ is the averaged concentration in cell $(i, j)$. However, this formula assumes a uniform surfactant distribution along the cell, which may result in a significant inaccuracy, as previously noted in \cite{james_surfactant-conserving_2004} and confirmed by our numerical experiments. To mitigate this issue, it was proposed in \cite{james_surfactant-conserving_2004} to redistribute the surfactant concentration along the reconstructed interface in a preliminary step. To this end, let us consider Fig.~\ref{fig:area-flux-estimation}(\textit{c}), and let $s$ denote the arc-length parameter along the reconstructed interface within cell $(i, j)$, with $s = 0$ and $s = \widetilde A^{n - 1/2}_i$ at the left and right boundaries, respectively. Assuming a constant surfactant concentration gradient $(\nabla_s\mathit{\Gamma})^{n - 1/2}_i$, \textit{i.e.} a linear surfactant distribution along the cell, the mass flux $\psi^{n - 1/2}_{i+1/2}$ across the cell face $i+1/2$ can be estimated as
\begin{align}\label{eq:16}
  \psi^{n - \frac12}_{i+\frac12} &= \int_{\widetilde A^{n - \frac12}_i - \phi^{n - \frac12}_{i+\frac12}}^{\widetilde A^{n - \frac12}_i} \bigg ((\nabla_s\mathit{\Gamma})_i^{n - \frac12}\left(s-\frac{\widetilde A^{n - \frac12}_i}{2}\right) + \mathit{\Gamma}^{n - \frac12}_i \bigg )~\mathrm{d}\ell \nonumber \\
                                 &=\frac12(\nabla_s\mathit{\Gamma})^{n - \frac12}_i \bigg (\widetilde A^{n - \frac12}_i\phi^{n - \frac12}_{i+\frac12} - (\phi^{n - \frac12}_{i+\frac12})^2 \bigg) + \mathit{\Gamma}^{n - \frac12}_i\phi^{n - \frac12}_{i+\frac12}.
\end{align}

Alternatively, if $\phi^{n - 1/2}_{i+1/2}$ is negative, the interface is advected from cell $i+1$ to cell $i$ and a similar procedure is applied to estimate the negative mass flux $\psi^{n - 1/2}_{i+1/2}$, with integration performed over the reconstructed interface in cell $i+1$ to ensure numerical consistency. Extensions of this approach to axisymmetric and 3D configurations are detailed in \ref{sec:mass-flux-calc} and \ref{sec:mass-flux-calc-1}, respectively.

\subsubsection{Surfactant concentration update in the advection step}\label{sec:update-surf-conc}

After  the mass fluxes $\psi^{n - 1/2}_{i-1/2}$ and $\psi^{n - 1/2}_{i+1/2}$ at the left and right faces of cell $(i, j)$ are obtained, the surfactant mass in cell $i$ can be updated using Eq.~(\ref{eq:aa3}). In this $x$-direction sub-step, this yields
\begin{equation}\label{eq:17}
  \widehat{M}_i^* = M^{n-\frac12}_i + \psi^{n-\frac12}_{i-\frac12} - \psi^{n-\frac12}_{i+\frac12}\,.
\end{equation}

To compute the surfactant concentration $\mathit{\Gamma}^{*}_i = \widehat{M}_i^*/ A_i^*$, the interface area $A_i^*$ corresponding to cell $(i, j)$ is required. Setting ${A}^*_i = \widetilde A^*_i$, with $\widetilde A^*_i$ given by Eq. \eqref{eq:area-diff-2}, (hereinafter referred to as option I) obviously conserves the surfactant mass. 
However, this choice is not the most appropriate for computing the surfactant concentration, owing to the inaccuracies introduced in the evaluation of the interface area by compression/stretching effects. Since the surfactant concentration has a direct influence on the Marangoni force, we privilege the approximation $A^*_i=\widetilde A^*_i - \widetilde{\varepsilon}_i^{n-\frac12}$, with $\widetilde{\varepsilon}_i^{n-\frac12}$ given by Eq. \eqref{eq:14} (option II). Therefore, at the end of the advection sub-step in the $x$ direction, the surfactant concentration is updated as
\begin{equation}\label{eq:18}
  \mathit{\Gamma}^*_i = \frac{\widehat{M}_i^*}{A^*_i} =\frac{\widehat{M}_i^*}{\widetilde A^*_i - \widetilde{\varepsilon}_i^{n-\frac12}},
\end{equation}

Typical results provided by both options are illustrated in the test case presented in \ref{sec:comp-betw-opti}. In this test, a circle with an initially uniform surfactant concentration is translating with a constant velocity. As the results show, the calculated surfactant concentration based on Eq. \eqref{eq:18} (option II) remains constant, as expected in this configuration which does not involve any stretching or compression of the interface. In contrast, predictions for $\mathit{\Gamma}$ based on the straightforward option I exhibit significant variations along the interface. Since option II does not strictly ensure conservation of the surfactant mass, we follow Xu \textit{et al.} \cite{xu_level-set_2012} and Teigen \textit{et al.} \cite{erik_teigen_diffuse-interface_2011} to mitigate this drawback. That is, we apply a rescaling procedure in the form $\mathit{\Gamma}^* = \beta\mathit{\Gamma}^*$ after each unidirectional advection step. The rescaling factor, $\beta$, is defined as the ratio of the total surfactant mass over the whole interface before and after the advection step. Numerical tests indicate that $\beta$ stays very close to $1$ with the relative error on the total mass decreasing to $\mathcal{O}(10^{-4})$ as the mesh is refined. For instance, in the test case reported in Fig. \ref{fig:shear_flow_2D_cov}, $|\beta-1|$ ranges from $5\times10^{-3}$ to  $8\times10^{-4}$ when the resolution is increased by a factor of four.

After the advection equation \eqref{eq:aa1} has been solved along the $x$ direction and $\mathit{\Gamma}^*$ is obtained, the same procedure is applied in the $y$ direction, through the sequence $\phi^*$ $\rightarrow$ $\psi^*$ $\rightarrow$ $\widehat M^{n+1/2}$ $\rightarrow$ $M^{n+1/2}$, yielding the intermediate concentration $\mathit{\Gamma}^{**}$. The extension to 3D is straightforward, requiring only a similar additional advection step in the $z$ direction.

\subsection{Sharp method for the diffusion step}\label{sec:geom-volume-fram}

Solving the diffusion step (\ref{eq:aa2}) provides the final update to the surfactant concentration, which transits from $ \mathit{\Gamma}^{**}$ to $\mathit{\Gamma}^{n+1/2}$ at the end of the time step $(n+1/2)$. To preserve second-order temporal accuracy in the overall method, the semi-implicit Crank - Nicolson scheme is employed for time discretization. The discretized form of (\ref{eq:aa2}) is expressed as
\begin{equation}\label{eq:21}
  \frac{\mathit{\Gamma}^{n+\frac12}-\mathit{\Gamma}^{**}}{\Delta t} = \frac12\left(\nabla_s\cdot(D_s\nabla_s\mathit{\Gamma}^{n+\frac12})+ \nabla_s\cdot(D_s\nabla_s\mathit{\Gamma}^{**})\right)  \qquad \forall\bm{x} \in \Upsilon.
\end{equation}

The spatial discretization of Eq.~(\ref{eq:21}) relies on the geometric information provided by the reconstructed interface. Given the substantial differences in the design of the numerical method in 2D and 3D configurations, the two cases are discussed separately in \S~\ref{sec:two-dimensional-case} and \S~\ref{sec:three-dimens-case}, respectively.

\subsubsection{Diffusion step in 2D configurations}\label{sec:two-dimensional-case}
\begin{figure}
  \centering
  \includegraphics[width=.8\textwidth]{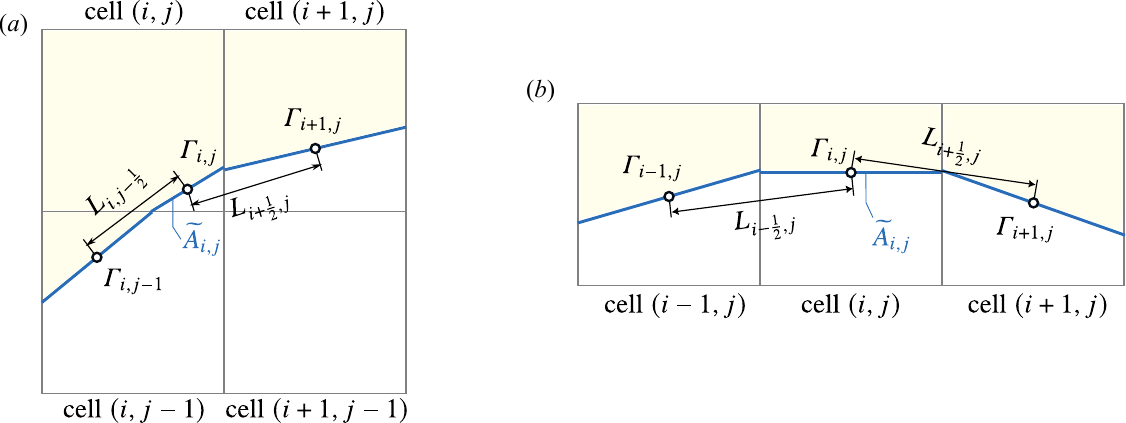}
  \caption{Illustration of the 2D spatial discretization used for solving the surfactant diffusion equation. (\textit{a}) and (\textit{b}) display the two possible types of neighborhood for an interfacial cell $(i,j)$.}
  \label{fig:2d-diffusion}
\end{figure}

The 2D reconstructed interface consists of a succession of line segments, each of which lies in an interfacial cell. Each segment is treated as an elementary control volume, with diffusion only taking place through its ends.  Fig.~\ref{fig:2d-diffusion} illustrates this sharp-interface approach. Integrating Eq.~(\ref{eq:21}) both spatially over the cell $(i,j)$ and over time yields
\begin{equation}\label{eq:22}
  \int_{\Delta t} \int_{\Upsilon_{i,j}^{n+\frac12}} \frac{\mathit{\Gamma}^{n+\frac12}-\mathit{\Gamma}^{**}}{\Delta t}~\mathrm{d}\ell~\mathrm{d}t = \int_{\Delta t} \int_{\Upsilon_{i,j}^{n+\frac12}}\frac12\left(\nabla_s\cdot(D_s\nabla_s\mathit{\Gamma}^{n+\frac12})+ \nabla_s\cdot(D_s\nabla_s\mathit{\Gamma}^{**})\right)~\mathrm{d}\ell~\mathrm{d}t\,,
\end{equation}
where $\Upsilon_{i,j}^{n+\frac12}$ represents the interface segment contained within cell $(i,j)$ at time step $(n+1/2)$. The left-hand side of Eq.~(\ref{eq:22}) can be approximated by using the concentration value at the centroid of the reconstructed interface, yielding
\begin{equation}\label{eq:23}
  \int_{\Delta t} \int_{\Upsilon_{i,j}^{n+\frac12}} \frac{\mathit{\Gamma}^{n+\frac12}-\mathit{\Gamma}^{**}}{\Delta t}~\mathrm{d}\ell~\mathrm{d}t \approx  \widetilde A^{n+\frac12}_{i,j} \left (\mathit{\Gamma}^{n+\frac12}_{i,j} - \mathit{\Gamma}^{**}_{i,j}\right),
\end{equation}
where $\widetilde A^{n+1/2}_{i,j}$ is the area of the reconstructed interface $\Upsilon_{i,j}^{n+1/2}$. The right-hand side of Eq.~(\ref{eq:22}) can be rewritten as the sum of the surfactant fluxes entering and leaving the control volume. For simplicity, the superscripts $(n+1/2)$ and $*$ are omitted, and the time-space integration is reformulated as
\begin{equation}\label{eq:24}
  \int_{\Delta t} \int_{\Upsilon_{i,j}^{n+\frac12}} \nabla_s\cdot(D_s\nabla_s\mathit{\Gamma}) ~\mathrm{d}\ell~\mathrm{d}t = \Delta t \sum (D_s\nabla_s\mathit{\Gamma})_f \cdot\bm{\tau}_f \approx
  F_{i+\frac12,j} + F_{i-\frac12,j} + F_{i,j+\frac12} + F_{i,j-\frac12}\,,
\end{equation}
where $\sum$ denotes summation over the cell faces, and $\bm{\tau}_f$ is the unit tangent vector lying along the interface $\Upsilon_{i,j}$, pointing outward the cell. Additionally, $F_{i\pm1/2,j}$ and $F_{i,j\pm1/2}$ represent the surfactant flux across the respective cell faces, with positive and negative values corresponding to outgoing and incoming fluxes, respectively. It is important to note that each 2D interfacial cell has only two neighboring interfacial cells, resulting in only two nonzero fluxes $F$. Figures \ref{fig:2d-diffusion}(\textit{a}-\textit{b}) depict two potential geometric configurations for the neighborhood of an interfacial cell $(i,j)$. In panel (\textit{a}), the neighboring cells are diagonally positioned (they belong to two different rows), leading to $F_{i-1/2,j} = F_{i,j+1/2} = 0$, while in panel (\textit{b}) they lie on the same row, implying $F_{i,j+1/2} = F_{i,j-1/2} = 0$. The surfactant mass flux in Eq.~(\ref{eq:24}) can then be approximated as
\begin{align}\label{eq:25}
  F_{i+\frac12,j} \approx D_{s(i+\frac12,j)} \Delta t \frac{\mathit{\Gamma}_{i+1,j} - \mathit{\Gamma}_{i,j}}{L_{i+\frac12,j}},
  \qquad &
           F_{i-\frac12,j} \approx D_{s(i-\frac12,j)}  \Delta t \frac{\mathit{\Gamma}_{i-1,j} - \mathit{\Gamma}_{i,j}}{L_{i-\frac12,j}}, \nonumber\\
  F_{i,j+\frac12} \approx D_{s(i,j+\frac12)}  \Delta t \frac{\mathit{\Gamma}_{i,j+1} - \mathit{\Gamma}_{i,j}}{L_{i,j+\frac12}},
  \qquad &
           F_{i,j-\frac12} \approx D_{s(i,j-\frac12)}  \Delta t \frac{\mathit{\Gamma}_{i,j-1} - \mathit{\Gamma}_{i,j}}{L_{i,j-\frac12}},
\end{align}
where $D_{s(i\pm1/2,j)}$ and $D_{s(i,j\pm1/2)}$ denote the surface diffusivities at the corresponding cell faces, and $L_{i\pm1/2,j}$ and $L_{i,j\pm1/2}$ represent the distances from the interface centroid in cell $(i,j)$ to that in the relevant neighboring cell, as illustrated in Figs.~\ref{fig:2d-diffusion}(\textit{a}-\textit{b}).

Substituting Eqs.~(\ref{eq:23}) and (\ref{eq:25}) into Eq.~(\ref{eq:21}) results in an implicit linear system for $\mathit{\Gamma}_{i,j}^{n+1/2}$. This system can be solved efficiently by using the multi-grid iterative approach implemented in \textit{Basilisk} \cite{popinet_gerris_2003}. After the solution is obtained, the update of the surfactant concentration to the next time step, $(n+1/2)$, is completed in the 2D case.

\subsubsection{Diffusion step in 3D configurations}\label{sec:three-dimens-case}

\begin{figure}
  \centering
  \includegraphics[width=\textwidth]{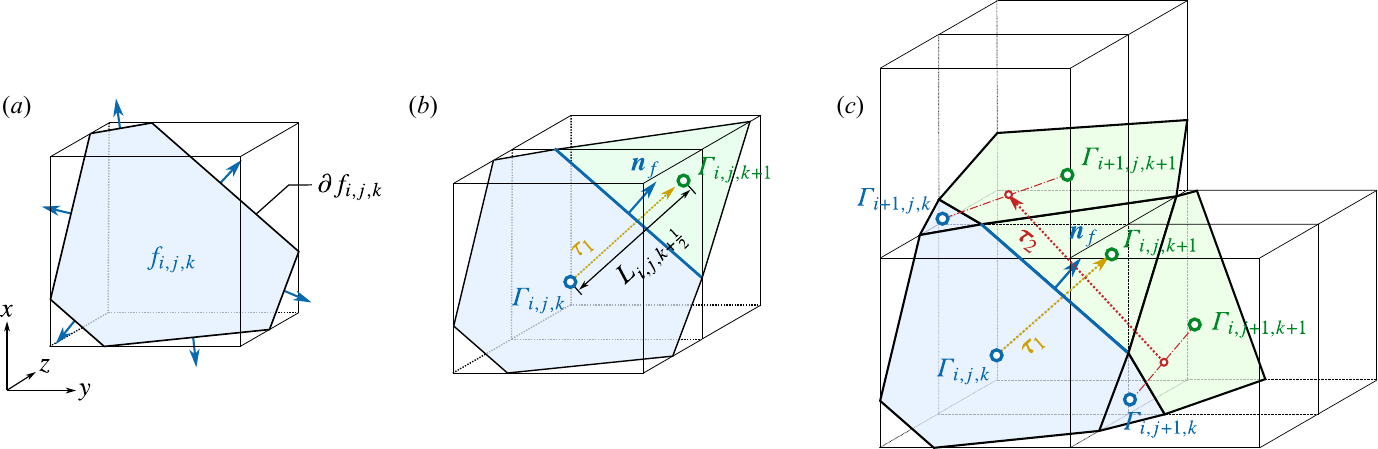}
    \vspace{-3mm}
  \caption{Illustration of the 3D spatial discretization for solving the surfactant diffusion equation. (\textit{a}) The reconstructed interface $f_{i,j,k}$ in the cell $(i,j,k)$ is regarded as the control volume; (\textit{b}) simple method for calculating the surfactant flux $F_{i,j,k+1/2}$; (\textit{c}) higher-order method for calculating $F_{i,j,k+1/2}$ with the aid of the surface gradient in two different tangential directions, $\bm{\tau}_1$ and $\bm{\tau}_2$.}
  \label{fig:3d-diffusion}
\end{figure}

Overall, the 3D methodology is similar to that detailed in the 2D case. However, interpolation schemes are more complex due to the interface within an interfacial cell being now a facet rather than a line segment, which offers numerous topological combinations with the neighboring interfacial cells. Figure \ref{fig:3d-diffusion}(\textit{a}) depicts the facet corresponding to the reconstructed interface lying in the interfacial cell $(i, j, k)$. The time-space integration of Eq.~(\ref{eq:21}) now yields
\begin{equation}\label{eq:26}
  \int_{\Delta t} \int_{\Upsilon_{i,j,k}^{n+\frac12}}\frac{\mathit{\Gamma}^{n+\frac12} - \mathit{\Gamma}^{**}}{\Delta t}~\mathrm{d}\mathcal{A}~\mathrm{d}t =  \int_{\Delta t}  \int_{\Upsilon_{i,j,k}^{n+\frac12}} \frac12\left(\nabla_s\cdot(D_s\nabla_s\mathit{\Gamma}^{n+\frac12})+ \nabla_s\cdot(D_s\nabla_s\mathit{\Gamma}^{**})\right)~\mathrm{d}\mathcal{A}~\mathrm{d}t\,,
\end{equation}
with $\Upsilon_{i,j,k}^{n+1/2}$ denoting the interface facet in cell $(i,j,k)$. The left-hand side of (\ref{eq:26}) can still be approximated using the surfactant concentration at the centroid of the reconstructed interface, so that
\begin{equation}\label{eq:27}
  \int_{\Delta t}   \int_{\Upsilon_{i,j,k}^{n+\frac12}}\frac{\mathit{\Gamma}^{n+\frac12} - \mathit{\Gamma}^{**}}{\Delta t}~\mathrm{d}\mathcal{A}~\mathrm{d}t \approx \widetilde A^{n+\frac12}_{i,j,k} \left(\mathit{\Gamma}^{n+\frac12}_{i,j,k} - \mathit{\Gamma}^{**}_{i,j,k} \right )\,,
\end{equation}
where $\widetilde A^{n+1/2}_{i,j,k}$ represents the area of the interface facet in cell $(i,j,k)$. Using the divergence theorem, the right-hand side of (\ref{eq:26}) may be expressed as the sum of the surfactant fluxes across the edges of $\Upsilon_{i,j,k}$, as illustrated by the blue arrows in Fig.~\ref{fig:3d-diffusion}(\textit{a}). Omitting the superscript $n+\frac{1}{2}$, the total flux can be approximated as
\begin{equation}\label{eq:28}
  \int_{\Delta t} \int_{\Upsilon_{i,j,k}}\bigg(\nabla_s\cdot(D_s\nabla_s\mathit{\Gamma})\bigg)~\mathrm{d}\mathcal{A} ~\mathrm{d}t\approx \Delta t\sum (D_s\nabla_s\mathit{\Gamma})_f\cdot\bm{n}_f~d_f \approx F_{i+\frac12,j,k} + F_{i-\frac12,j,k} + F_{i,j+\frac12,k} + F_{i,j-\frac12,k} + F_{i,j,k+\frac12} + F_{i,j,k-\frac12}\,.\,\,
\end{equation}
In \eqref{eq:28}, $\sum$ denotes the summation over the six cell faces, $d_f$ represents the length of the considered edge (thick blue line in Fig.~\ref{fig:3d-diffusion}(\textit{b})) and $\bm{n}_f$ is the outward unit normal vector at the edge of the facet, lying in the plane of the interface. Terms $F_{i\pm1/2, j, k}$, $F_{i, j\pm1/2, k}$, and $F_{i, j, k\pm1/2}$ represent the surfactant flux across the corresponding cell face. Unlike the 2D case, where the flux is nonzero only across two cell faces, the number of neighboring interfacial cells surrounding a given cell in 3D varies according to the local configuration. Figure~\ref{fig:3d-diffusion}(\textit{a}) presents a scenario where all six flux components are nonzero. The remaining challenge is to estimate the non-zero flux $F$ at the cell faces. The most straightforward approach is to use a two-point finite-difference discretization, similar to the 2D case. For instance, considering the cell face $(i, j, k+1/2)$ in Fig.~\ref{fig:3d-diffusion}(\textit{b}), the flux can be approximated as
\begin{equation}\label{eq:30}
  F_{i,j,k+\frac12} \approx d_{f(i,j,k+\frac12)}D_{s(i,j,k+\frac12)}\Delta t \frac{\mathit{\Gamma}_{i,j,k+1} - \mathit{\Gamma}_{i,j,k}}{L_{i,j,k+\frac12}}=d_{f(i,j,k+\frac12)}D_{s(i,j,k+\frac12)}\Delta t \,g_1\,.
\end{equation}
However, Eq.~(\ref{eq:30}) actually defines the flux along the line connecting the two cell centroids, characterized by a vector $\bm{\tau}_1$ with length $L_{i, j, k+1/2}$, as shown in Fig.~\ref{fig:3d-diffusion}(\textit{b}). Since the two cells are not necessarily coplanar, $\bm{\tau}_1$ is generally not parallel to $\bm{n}_f$ and the derivative $g_1$ evaluated in the right-hand side of \eqref{eq:30} is actually the concentration gradient along $\bm{\tau}_1$, \textit{i.e.} $ g_1 = (\nabla_{s}\mathit{\Gamma})_{i, j, k+\frac12}\cdot\bm{\tau}_1$. If $\bm{n}_f$ and $\bm{\tau}_1$ are not parallel, $g_1$ does not converge to the expected projection $ (\nabla_{s}\mathit{\Gamma})_{i, j, k+\frac12}\cdot\bm{n}_f$ as the spatial resolution increases, making the simple interpolation scheme defined by (\ref{eq:30}) inaccurate.

Consequently, a higher-order interpolation scheme is required. 
To this end, we need to determine a second vector $\bm{\tau}_2$, non collinear with $\bm{\tau}_1$, on which the concentration gradient may also be projected.  
For this purpose, we extend the cell template by including four additional interfacial cells as sketched in Fig.~\ref{fig:3d-diffusion}(\textit{c}), namely, cells $(i+1, j, k)$, $(i+1,j,k+1)$, $(i,j+1,k)$, and $(i,j+1,k+1)$. We then connect the centroids of the interface facets contained in cells $(i+1, j, k)$ and $(i+1,j,k+1)$, as well as those contained in cells $(i,j+1,k)$, and $(i,j+1,k+1)$, as indicated by the two red dashed-dotted lines in Fig.~\ref{fig:3d-diffusion}(\textit{c}). Next, we define $\bm{\tau}_2$ as the vector that connects the midpoints of these two lines, the length of which is  $L'_{i,j,k+1/2}$. The  concentration gradient along $\bm{\tau}_2$ can then be estimated as
\begin{equation}\label{eq:32}
  g_2 = (\nabla_{s}\mathit{\Gamma})_{i, j, k+\frac12}\cdot\bm{\tau}_2 \approx \frac{\mathit{\Gamma}_{i+1,j,k}+\mathit{\Gamma}_{i+1,j,k+1}-\mathit{\Gamma}_{i,j+1,k}-\mathit{\Gamma}_{i,j+1,k+1}}{2L'_{i,j,k+\frac12}}\,.
\end{equation}

Since $\bm{\tau}_1$ and $\bm{\tau}_2$ are not necessarily perpendicular, we compute a third tangential vector $\bm{\tau}_3 = \bm{\tau}_1\times (\bm{n}_{i,j,k}+\bm{n}_{i,j,k+1})/2$, obviously normal to $\bm{\tau}_1$, with $\bm{n}_{i,j,k}$ and $\bm{n}_{i,j,k+1}$ the unit normals to the reconstructed interface in cells $(i,j,k)$ and $(i,j,k+1)$, respectively. The concentration surface gradient can now be expressed as
\begin{equation}\label{eq:33}
  (\nabla_s\mathit{\Gamma})_{i, j, k+\frac12} = g_1\bm{\tau}_1+g_3\bm{\tau}_3\,,
\end{equation}
where $g_3$ may be obtained by projecting (\ref{eq:32}) onto $\bm{\tau}_2$ and making use of (\ref{eq:33}), yielding
\begin{equation}\label{eq:34}
  g_3 = \frac{g_2 - g_1\bm{\tau}_1\cdot\bm{\tau}_2}{\bm{\tau}_2\cdot\bm{\tau}_3}\,.
\end{equation}
Finally, instead of the basic estimate \eqref{eq:30}, \eqref{eq:33} and \eqref{eq:34} allow the flux $F_{i,j,k+1/2}$ to be estimated as
\begin{align}\label{eq:35}
  F_{i,j,k+\frac12} \approx 
                     d_{f(i,j,k+\frac12)}D_{s(i,j,k+\frac12)}\Delta t \left(g_1\bm{\tau}_1\cdot\bm{n}_f + \frac{g_2 - g_1\bm{\tau}_1\cdot\bm{\tau}_2}{\bm{\tau}_2\cdot\bm{\tau}_3} \bm{\tau}_3\cdot\bm{n}_f\right)\,,
\end{align}
with $g_1$ and $g_2$ computed from (\ref{eq:30}) and (\ref{eq:32}), respectively. A similar procedure may be applied to every face of an interfacial cell.
The advantages of the above approximation over the basic approximation (\ref{eq:30}) will be demonstrated on a 3D diffusion test case in \S~\ref{sec:three-dimens-case-1}.
The combination of Eqs.~(\ref{eq:27}-\ref{eq:32}), and (\ref{eq:35}) constitutes the space-time discretized system for the diffusion step of the surfactant concentration equation in 3D, which is still solved using the multi-grid iterative approach implemented in \textit{Basilisk}. As with the 2D case, this step completes the evolution of the surfactant concentration from the previous time step $(n - 1/2)$ to the next, $(n+1/2)$.

\subsection{Overall time-advancement sequence}\label{sec:over-disr-steps}

\begin{figure}[h]
  \centering
  \includegraphics[width=0.95\textwidth]{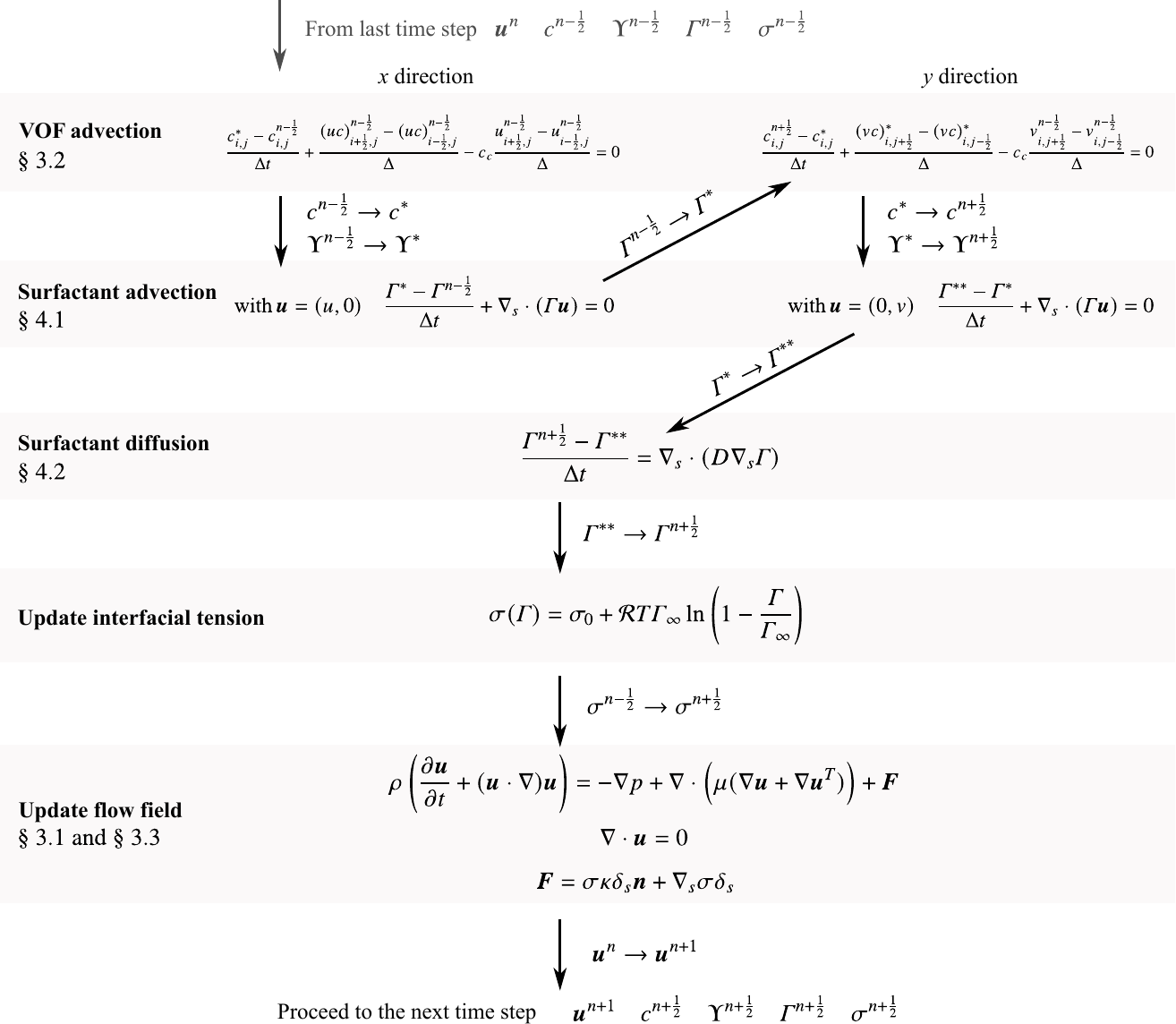}
  \caption{Schematic representation of the numerical steps involved in the transport of the surfactant concentration over one time step for a 2D case. The extension to 3D is straightforward. The succession of arrows indicates the order in which the various steps are executed.}
  \label{fig:global-procedures-whole-schema}
\end{figure}

To sum up, the overall sequence implemented in \textit{Basilisk} with the present sharp interface method to solve the surfactant transport equation and, beyond it, the coupled two-phase flow problem, is schematically illustrated in Fig.~\ref{fig:global-procedures-whole-schema}. The illustration is for the 2D case but its extension to the 3D case is straightforward.

\section{Numerical tests}\label{sec:validation-cases}

In this section, the proposed numerical method is assessed and validated through a series of benchmark test cases. First, individual modules, such as those in charge of the advection and diffusion of the surfactant and the computation of the Marangoni force, are assessed in \S~\ref{sec:surfactant-advection} to \S~\ref{sec:marang-force-calc-1} by comparing the corresponding predictions with those of previous numerical studies. Subsequently, these modules are coupled to simulate interfacial flows in the presence of insoluble surfactants. Results for a 2D case are presented in \S~\ref{sec:2d-surfactant-laden}, followed by those in axisymmetric and 3D configurations in \S~\ref{sec:dropl-deform-an}.

\subsection{Surfactant advection}\label{sec:surfactant-advection}

This test case, originally proposed by Cleret de Langavant \textit{et al.} \cite{cleret_de_langavant_level-set_2017}, evaluates the accuracy of numerical methods in the advection of insoluble surfactants. A 2D or 3D star-shaped interface is initially laden uniformly with a concentration $\mathit{\Gamma}_{t=0} = 0.5$. Under a prescribed non-solenoidal flow field, $\bm{u} = {\bm{x}}/{||\bm{x}||}$, the interface expands over time, inducing surfactant advection along the interface. This test case ensures a nonzero stretching term in Eq.~(\ref{eq:strech-compress}). Thus, it is well suited for evaluating the accuracy of the sharp-interface numerical scheme, especially the estimation of the interface area employed in the advection step (see \S\ref{sec:cons-advect-meth}). We analyze the transient distribution of the surfactant along the interface, presenting the 2D and 3D results in \S~\ref{sec:two-dimensional-case-2} and \S~\ref{sec:three-dimens-case-2}, respectively.

\subsubsection{Two-dimensional case}\label{sec:two-dimensional-case-2}

\begin{figure}
  \centering
  \includegraphics[width=\textwidth]{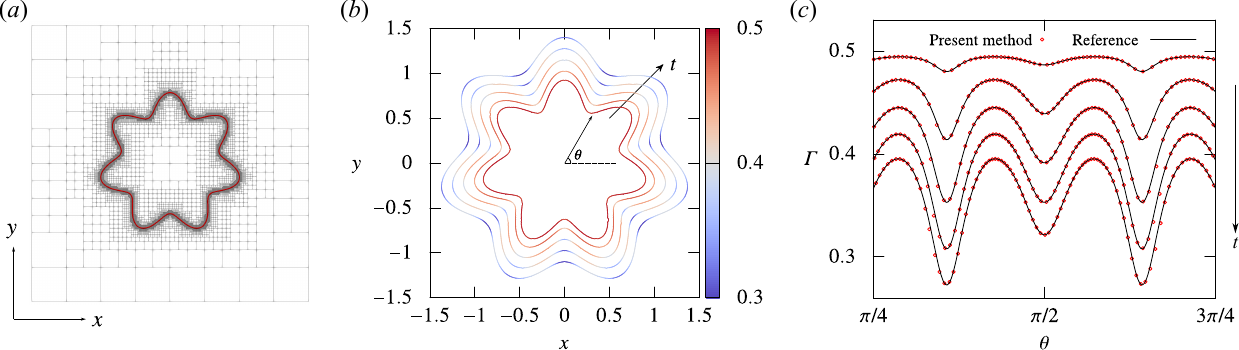}
    \vspace{-5mm}
  \caption{2D surfactant advection test. (\textit{a}) Snapshot of the reconstructed interface and adaptive grid at $t=0.125$; (\textit{b}) series of interfaces colored by surfactant concentration $\mathit{\Gamma}$ at $t=0.025,0.125,0.25,0.375,0.5$; (\textit{c}) surfactant concentration distribution along the interface (red circles) compared with the analytical solution \eqref{test2D} (black line) at $t=0.025,0.125,0.25,0.375,0.5$; the polar angle $\theta$ is defined in panel (\textit{b}).}
  \label{fig:star_2D}
\end{figure}

The computational domain is the $[-2,2]\times[-2,2]$ square, and the initial star-like interface is prescribed using the polar equation
\begin{equation}\label{eq:36}
  r(\theta, t = 0) = 0.75(1-0.2\sin 7\theta)\,,
\end{equation}
where $r(\theta)$ denotes the radial position as a function of the polar angle $\theta$, the definition of which is sketched in Fig.~\ref{fig:star_2D}(\textit{b}). From the conservation of surfactant mass, the analytical solution for the surfactant concentration along the evolving interface is given by
\begin{equation}
\label{test2D}
  \mathit{\Gamma}(\theta,t) = \frac12\frac{\sqrt{r^2(\theta)+\dot{r}_\theta^2(\theta)}}{\sqrt{(t+r(\theta))^2+\dot{r}_\theta^2(\theta)}},
\end{equation}
where $\dot{r}_\theta(\theta) = \mathrm{d}r/\mathrm{d}\theta$. The computational domain is discretized using the adaptive mesh refinement (AMR) strategy embedded in \textit{Basilisk}, with a resolution $1/\Delta_{\mathrm{min}}=128$ near the interface and $1/\Delta_{\mathrm{max}}=2$ in the far field. The time step is set to $\Delta t = 10^{-3}$, and the simulation runs until $t=0.5$.

The numerical results are presented in Fig.~\ref{fig:star_2D}. Panel (\textit{a}) illustrates the interface shape and computational grid at  time $t = 0.125$. The temporal evolution of the interface and the associated surfactant distribution are shown in Fig.~\ref{fig:star_2D}(\textit{b}), where the concentration is seen to decrease as the interface expands. Lower concentrations are observed at the tips and troughs of the star-shaped interface, while higher concentrations persist in intermediate regions. Despite the use of a Cartesian mesh, the concentration distribution remains rotationally symmetric, free from grid-induced anisotropy. Quantitative comparisons in Fig.~\ref{fig:star_2D}(\textit{c}) demonstrate excellent agreement between the numerical and analytical solutions.

\subsubsection{Three-dimensional case}\label{sec:three-dimens-case-2}

\begin{figure}
  \centering
  \includegraphics[width=\textwidth]{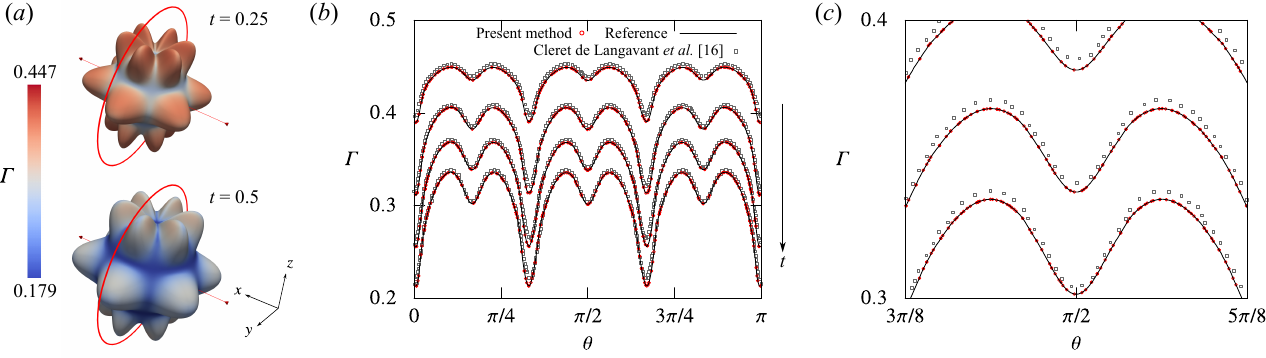}
    \vspace{-5mm}
  \caption{3D surfactant advection test. (\textit{a}) Interface snapshots colored by surfactant concentration at $t=0.25$ and $t=0.5$; (\textit{b}) surfactant concentration distribution along a plane passing through the star's apexes (red circles in panel (\textit{a})), compared with the analytical solution (black line) and numerical results from \cite{cleret_de_langavant_level-set_2017} (grey squares) at $t = 0.1, 0.2, 0.3, 0.4$ (from top to bottom); (\textit{c}) zoomed-in view of (\textit{b}).}
  \label{fig:star_3D}
\end{figure}

The 3D case is simulated using a cubic computational domain $[-2,2]\times[-2,2]\times[-2,2]$, where the initial star-like interface is defined in spherical coordinates as
\begin{equation}
  r(\theta,\varphi, t = 0) = 0.75+ 0.2(1 - 0.6\cos 6\varphi)(1- \cos 6 \theta)\,,
\end{equation}
where $r(\theta,\varphi)$ denotes the radial position of the interface as a function of the polar angle $\theta = \mathrm{cos}^{-1}(z/\sqrt{x^2+y^2+z^2})$ and the azimuthal angle $\varphi = \mathrm{sgn}(y)\,\mathrm{cos}^{-1}(x/\sqrt{x^2+y^2})$. The analytical solution for the transient surfactant concentration, obtained from mass conservation, is given by \cite{cleret_de_langavant_level-set_2017}
\begin{equation}
  \mathit{\Gamma}(\theta,\varphi,t) = \frac12\frac{\sqrt{r^2(\theta,\varphi)+\dot{r}_\theta^2(\theta,\varphi)}\sqrt{r^2(\theta,\varphi)+\dot{r}_\varphi^2(\theta,\varphi)}}{\sqrt{(t+r(\theta,\varphi))^2+\dot{r}_\theta^2(\theta,\varphi)}\sqrt{(t+r(\theta,\varphi))^2+\dot{r}_\varphi^2(\theta,\varphi)}}\,,
\end{equation}
where $\dot{r}_\theta$ and $\dot{r}_\varphi$ denote partial derivatives with respect to $\theta$ and $\varphi$, respectively. The AMR strategy refines cells near the interface up to a resolution $1/\Delta_{\mathrm{min}}=64$. The time step is still set to $\Delta t = 10^{-3}$, and the simulation proceeds until $t=0.5$.

The numerical results are shown in Fig.~\ref{fig:star_3D}. Panel (\textit{a}) displays the surfactant concentration distribution at $t=0.25$ and $t=0.5$. Similar to the 2D case, the concentration decreases as the interface expands, with lower concentrations observed in the trough regions. Figures \ref{fig:star_3D}(\textit{b}) and (\textit{c}) show the time evolution of the surfactant concentration distribution in a cross-sectional plane indicated with a red circle in panel (\textit{a}). 
Numerical results from \cite{cleret_de_langavant_level-set_2017} are included for comparison, showing that the present sharp interface method provides more accurate results than the diffusive interface approach at the same spatial resolution.

\subsubsection{Grid convergence analysis}
\label{sec:convergence-analysis}

\begin{figure}
  \centering
  \includegraphics[width=.4\textwidth]{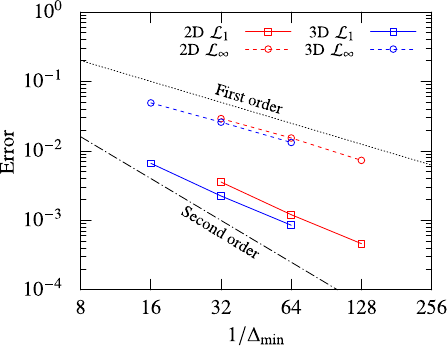}
  \caption{Convergence study of the $\mathcal{L}_1$ and $\mathcal{L}_\infty$ norms of the surfactant concentration error at $t=0.5$. Results are shown for both the 2D (\S~\ref{sec:two-dimensional-case-2}) and 3D (\S~\ref{sec:three-dimens-case-2}) advection tests.}
  \label{fig:adv-conv}
\end{figure}

To completely qualify the proposed numerical treatment of the surfactant advection term, it is of interest to examine the dependence of the computed concentration errors upon spatial resolution for the two test cases presented in \S\S~\ref{sec:two-dimensional-case-2} and \ref{sec:three-dimens-case-2}. Figure~\ref{fig:adv-conv} shows the $\mathcal{L}_1$ and $\mathcal{L}_\infty$ norms of these errors at $t=0.5$. The results indicate that the $\mathcal{L}_\infty$ norm exhibits a convergence close to first order, while the $\mathcal{L}_1$ norm converges at a faster rate,  halfway between first and second order. A more detailed analysis reveals that the maximum errors (captured by the $\mathcal{L}_\infty$ norm) consistently occur near the troughs of the interface, where the curvature is large. These errors primarily arise from the piecewise linear interface approximation on which the PLIC scheme is grounded, which is less accurate in small cells intersected by highly curved interfaces, as already noted by James and Lowengrub~\cite{james_surfactant-conserving_2004}. Incorporating higher-order curvature information through a local parabolic interface reconstruction would presumably improve accuracy in such regions, allowing a faster convergence of the proposed treatment of the surfactant advection flux.

\subsection{Surfactant diffusion}\label{sec:surf-diff-test}

In this subsection, the approach for solving the surfactant diffusion equation ((\ref{eq:aa2})) formulated in \S~\ref{sec:geom-volume-fram}, is validated. To facilitate a direct comparison with the diffuse interface method, we adopt the benchmark test case proposed in \cite{cleret_de_langavant_level-set_2017}. Given the distinct discretization schemes for two- and three-dimensional problems, we present the two tests separately.

\subsubsection{Two-dimensional case}\label{sec:two-dimensional-case-1}

\begin{figure}
  \centering
  \includegraphics[width=\textwidth]{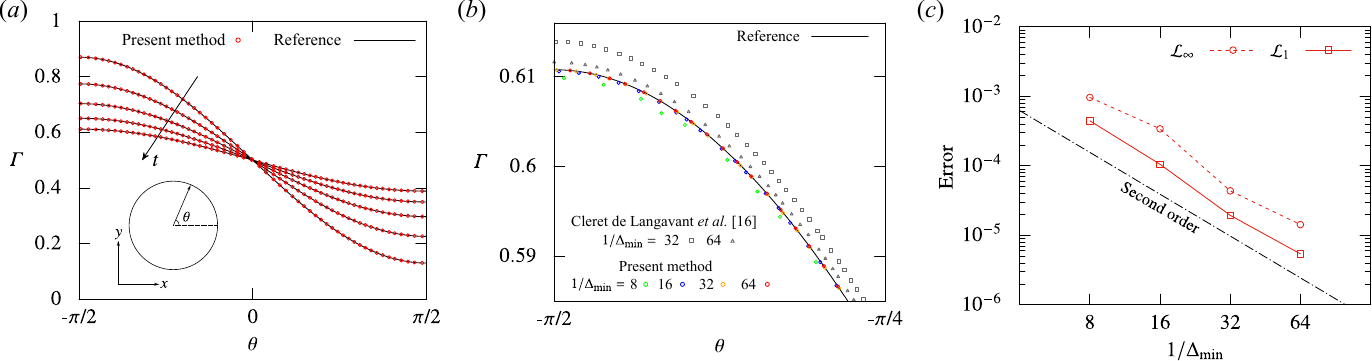}
    \vspace{-5mm}
  \caption{2D surfactant diffusion test. (\textit{a}) Evolution of the surfactant concentration along the interface at $t=0.3,0.6,0.9,1.2,1.5$, with $1/\Delta_{\mathrm{min}} = 64$, compared with the analytical solution \eqref{sol:diffusion}; (\textit{b}) comparison of the present numerical solution at $t=1.507$ with that of \cite{cleret_de_langavant_level-set_2017}, for different resolutions; (\textit{c}) grid convergence of the $\mathcal{L}_1$ and $\mathcal{L}_\infty$ norms of the surfactant concentration error at $t=1.507$.}
  \label{fig:diffusion_test_2D}
\end{figure}

We consider a stationary circular interface of radius $R=1$, centered at $(x,y)=(0,0)$, within a square computational domain $[-2,2]\times[-2,2]$. The initial surfactant concentration distribution is given by
\begin{equation}
  \mathit{\Gamma}(\theta,t=0) = \frac12\left(1+\sin\theta\right),
\end{equation}
where $\theta$ is the polar angle defined in Fig.~\ref{fig:diffusion_test_2D}(\textit{a}). In the absence of flow, surfactant transport is governed solely by diffusion, and the theoretical solution of the transport equation reads
\begin{equation}
\label{sol:diffusion}
  \mathit{\Gamma}(\theta,t) = \frac12\left(1+e^{-D_st/R^2}\sin\theta\right)\,.
\end{equation}
The numerical results obtained with the molecular diffusivity set to $D_s=1$ are presented in Fig.~\ref{fig:diffusion_test_2D}. Panel (\textit{a}) depicts the evolution of the surfactant concentration along the interface at different instants of time, with a spatial resolution $1/\Delta_{\mathrm{min}} = 64$ and a time step $\Delta t = 1.56\times10^{-2}$. The numerical results exhibit excellent agreement with the analytical solution, confirming the accuracy of the approach developed in \S~\ref{sec:geom-volume-fram} to solve the diffusion step on a sharp interface. A comparison between results provided by the present sharp interface method and the diffusive interface approach of Cleret de Langavant \textit{et al.} \cite{cleret_de_langavant_level-set_2017} is reported in Fig.~\ref{fig:diffusion_test_2D}(\textit{b}). This figure makes it clear that the present method achieves high accuracy even with the coarse resolution $1/\Delta_{\mathrm{min}} = 16$, demonstrating its superiority over the diffusive interface method at the same resolution. To further assess the numerical accuracy of the present approach, we performed a convergence analysis by varying the resolution from $1/\Delta_{\mathrm{min}} = 8$ to $1/\Delta_{\mathrm{min}} = 64$. The computed errors based on the $\mathcal{L}_1$ and $\mathcal{L}_\infty$ norms are shown in Fig.~\ref{fig:diffusion_test_2D}(\textit{c}), confirming that  the method is second-order accurate.
\begin{figure}
  \centering
  \includegraphics[width=\textwidth]{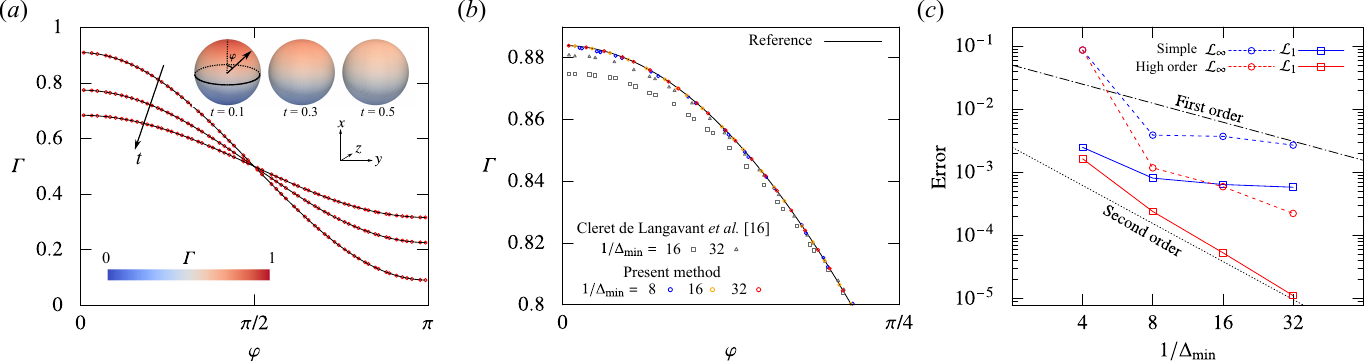}
    \vspace{-5mm}
  \caption{3D surfactant diffusion test. (\textit{a}) Surfactant concentration distribution in the diametrical plane $z= 0$ at $t=0.1,0.3,0.5$, with $1/\Delta_{\mathrm{min}} = 32$, compared with the analytical solution (the insets show the surfactant distribution over the whole interface in the half-plane $z<0$); (\textit{b}) comparison of concentration profiles in the plane $z=0$ at $t=0.132$ for different resolutions, using the present method and the approach of \cite{cleret_de_langavant_level-set_2017}; (\textit{c}) convergence analysis of the $\mathcal{L}_1$ and $\mathcal{L}_\infty$ norm errors at $t=0.132$, including a comparison with the simple flux computation method from Fig.~\ref{fig:3d-diffusion}(\textit{b}).}
  \label{fig:diffusion_test_3D}
\end{figure}

\subsubsection{Three-dimensional case}\label{sec:three-dimens-case-1}

We extend the analysis to 3D configurations by considering a stationary spherical interface of radius $R=1$, centered at $(x,y,z)=(0,0,0)$, within a cubic computational domain $[-2,2]\times[-2,2]\times[-2,2]$. The initial surfactant concentration distribution is prescribed as
\begin{equation}
  \mathit{\Gamma}(\varphi,t=0) = \frac12\left(1+\cos\varphi\right)\,,
\end{equation}
where $\varphi$ denotes the azimuthal angle $\varphi = \mathrm{sgn}(y)\,\mathrm{cos}^{-1}(x/\sqrt{x^2+y^2})$ (see Fig.~\ref{fig:diffusion_test_3D}(\textit{a})). The analytical solution is
\begin{equation}
  \mathit{\Gamma}(\varphi,t) = \frac12\left(1+e^{-2D_st/R^2}\cos\varphi\right)\,.
\end{equation}
The spatial resolution is increased from $1/\Delta_{\mathrm{min}} = 8$ up to $1/\Delta_{\mathrm{min}} = 32$; the time step reduces accordingly, from $\Delta t = 0.02$ down to $\Delta t = 2.5\times10^{-3}$, decreasing by a factor of two every time the spatial resolution is increased. Figure \ref{fig:diffusion_test_3D}(\textit{a}), which presents the evolution of the surfactant concentration along the interface at different time instants in the plane $z=0$, reveals an excellent agreement with the analytical solution. 
A direct comparison between the present sharp interface approach and the diffuse approach proposed in \cite{cleret_de_langavant_level-set_2017} is provided in Fig.~\ref{fig:diffusion_test_3D}(\textit{b}). Again, the results demonstrate that the sharp interface method achieves a superior accuracy at the same resolution. The convergence analysis based on the $\mathcal{L}_1$ and $\mathcal{L}_\infty$ norm errors at $t=0.132$ is reported in Fig.~\ref{fig:diffusion_test_3D}(\textit{c}) for both  the simple interpolation strategy corresponding to Eq.~(\ref{eq:30}) and the higher-order strategy (Eq.~(\ref{eq:35})). The results indicate that the accuracy of the former approach is significantly less than first order, while the higher-order interpolation achieves a convergence slightly better than first order in the $\mathcal{L}_\infty$ norm and a clear second-order convergence in the $\mathcal{L}_1$ norm. These findings further highlight the effectiveness of the proposed sharp-interface approach.

\subsection{Combined advection-diffusion in a 2D flow}\label{sec:2d-advect-diff}

\begin{figure}
  \centering
  \includegraphics[width=\textwidth]{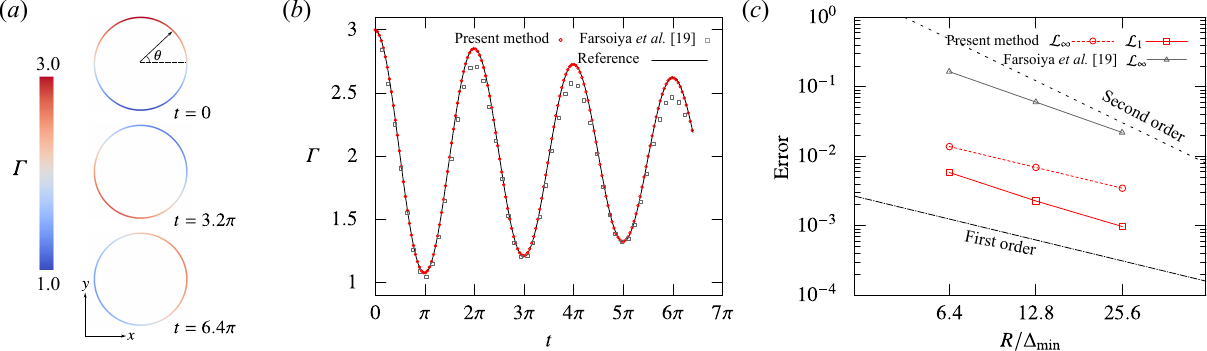}
    \vspace{-5mm}
  \caption{2D advection-diffusion test. (\textit{a}) Snapshots of the interface colored by the surfactant concentration at $t=0, 3.2\pi, 6.4\pi$ for $R/\Delta_{\mathrm{min}}=25.6$; (\textit{b}) time evolution of the surfactant concentration at position $\theta=\pi/2$ for $R/\Delta_{\mathrm{min}}=12.8$, compared with the analytical solution and the results of \cite{farsoiya_coupled_2024} at the same resolution; (\textit{c}) grid convergence of the $\mathcal{L}_1$ and $\mathcal{L}_\infty$ norms of the surfactant concentration error in the present approach and in that of \cite{farsoiya_coupled_2024}.}
  \label{fig:adv-diff}
\end{figure}
To close these series of tests on the advection and diffusion steps involved in the discretization of the surfactant concentration equation, we consider the configuration employed by Farsoiya \textit{et al.}~\cite{farsoiya_coupled_2024}, where surfactant transport takes place along a stationary 2D circular interface under a prescribed velocity field. The interface, a circle of radius $R=0.2$, is initially covered with the surfactant concentration
\begin{equation}
  \mathit{\Gamma}(\theta,t=0) = 2+\sin \theta\,,
\end{equation}
where $\theta$ is the polar angle (see Fig.~\ref{fig:adv-diff}(\textit{a})). The velocity field has only an azimuthal component, given by $\bm{u} = r\bm{e}_\theta$ with $r$ the radial position, ensuring that the interface remains stationary without deforming.

Simulations are performed in a domain $[-0.5,0.5]\times[-0.5,0.5]$,  the analytical solution for the evolution of the surfactant concentration being
\begin{equation}
  \mathit{\Gamma}(\theta,t) = 2+\sin(t+\theta)e^{-D_st/R^2},
\end{equation}
Figure \ref{fig:adv-diff} summarizes the numerical results obtained with the diffusivity set to $D_s=10^{-3}$ and $c_{\mathrm{CFL}}$ maintained at $0.5$. Panel (\textit{a}) presents snapshots of the interface, colored by the surfactant concentration, at three different instants of time for a fixed resolution $R/\Delta_{\mathrm{min}} = 12.8$.  The time evolution of the surfactant concentration at the interface apex displayed in panel \ref{fig:adv-diff}(\textit{b}) demonstrates excellent agreement with the analytical solution. The sharp interface method is seen to achieve a significantly higher accuracy than the diffuse interface approach of \cite{farsoiya_coupled_2024}, most notably around every maximum of the $ \mathit{\Gamma}(t)$ evolution. The grid  convergence analysis of the relative concentration error at the final time reported in panel \ref{fig:adv-diff}(\textit{c}) confirms the first-order accuracy of the overall approach (a slightly better convergence is achieved on the $\mathcal{L}_1$ norm). Additionally, the present method exhibits significantly lower errors, typically one order of magnitude smaller, compared with those produced by the diffuse interface method. As Sec. \ref{sec:convergence-analysis} made clear, the limitation of the overall convergence rate to first order results from the advection step, more precisely from limitations inherent to the PLIC interface reconstruction scheme.

\subsection{Marangoni force calculation}\label{sec:marang-force-calc-1}

\begin{figure}
  \centering
  \includegraphics[width=\textwidth]{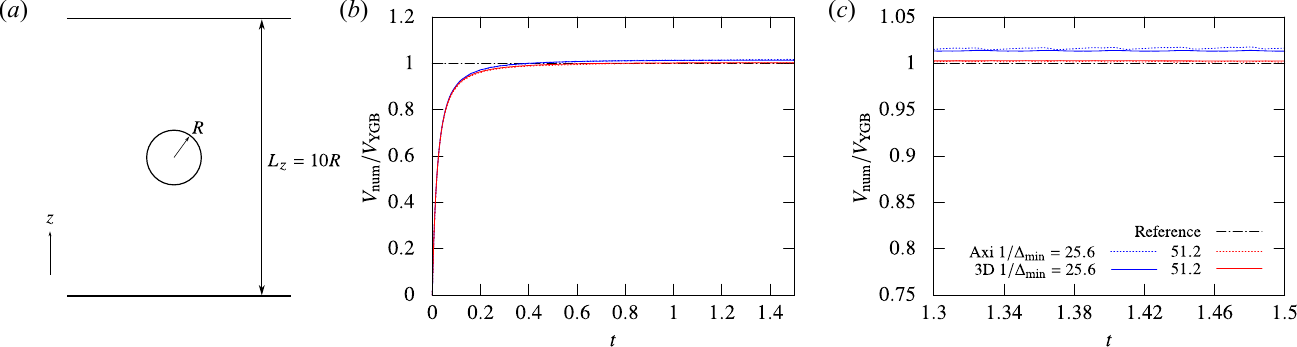}
  \vspace{-5mm}
  \caption{Marangoni force calculation test. (\textit{a}) Sketch of the configuration;
(\textit{b}) time evolution of the normalized droplet speed $V_{\mathrm{num}}/V_{\mathrm{YGB}}$; (\textit{c}) zoomed-in view of (\textit{b}) near the steady-state regime.
Results are displayed for both axisymmetric and 3D configurations at resolutions $1/\Delta_{\mathrm{min}} = 25.6$ and $1/\Delta_{\mathrm{min}} = 51.2$.}
  \label{fig:marangoni}
\end{figure}

To validate the Marangoni force calculation strategy described in \S~\ref{sec:marang-force-calc}, we adopt the test case proposed by Muradoglu and Tryggvason~\cite{muradoglu_simulations_2014}. This test, performed in both 2D axisymmetric and fully 3D configurations, considers the migration of a viscous droplet driven by Marangoni forces arising from an inhomogeneous surfactant distribution. The viscous droplet has a radius $R = 0.5$, and its center is positioned $5R$ from the domain boundaries to minimize boundary effects, as illustrated in Fig.~\ref{fig:marangoni}(\textit{a}). Instead of solving the surfactant transport equation, we directly impose a linear surfactant concentration profile along the domain's vertical axis $z$ in the form
 $\mathit{\Gamma}(z)/\mathit{\Gamma}_\infty = z/L_z$,
where $L_z = 10R$ is the height of the domain. A linear EOS $ \sigma(z) = \sigma_0\left(1-\beta_s{\mathit{\Gamma}(z)/{\mathit{\Gamma}_\infty}}\right)$ is employed instead of the nonlinear form in Eq.~(\ref{eq:eos}). The resulting hydrodynamic behavior is analogous to the thermocapillary migration of a viscous droplet in a fluid with a constant temperature gradient at vanishing Reynolds number, as analyzed by Young \textit{et al.} \cite{young_motion_1959}. At steady state, the theoretical migration speed of the droplet is
\begin{equation}
  V_{\mathrm{YGB}} = \frac{2\sigma_0\beta_sR}{L_z(6\mu_1+9\mu_2)}\,,
\end{equation}
where the viscosities and densities of each fluid are set to $\mu_1=\mu_2 = 0.1$ and $\rho_1=\rho_2=0.2$, with $\beta_s=2$ and $\sigma_0 = 1$.

Simulations are conducted at two spatial resolutions, $1/\Delta_{\mathrm{min}}=25.6$ and $1/\Delta_{\mathrm{min}}=51.2$, for both axisymmetric and 3D geometries. The time evolution of the normalized migration speed, $V_{\mathrm{num}}/V_{\mathrm{YGB}}$, is shown in Fig.~\ref{fig:marangoni}(b), with a closer view at the nearly steady state in Fig.~\ref{fig:marangoni}(c). The results exhibit excellent agreement with the theoretical prediction, and the speed evolutions are virtually identical in both geometries, demonstrating the robustness and self-consistency of the proposed numerical method for the Marangoni force calculation.

\subsection{Deformation of a 2D surfactant-laden drop in a shear flow}\label{sec:2d-surfactant-laden}

\begin{figure}
  \centering
  \includegraphics[width=\textwidth]{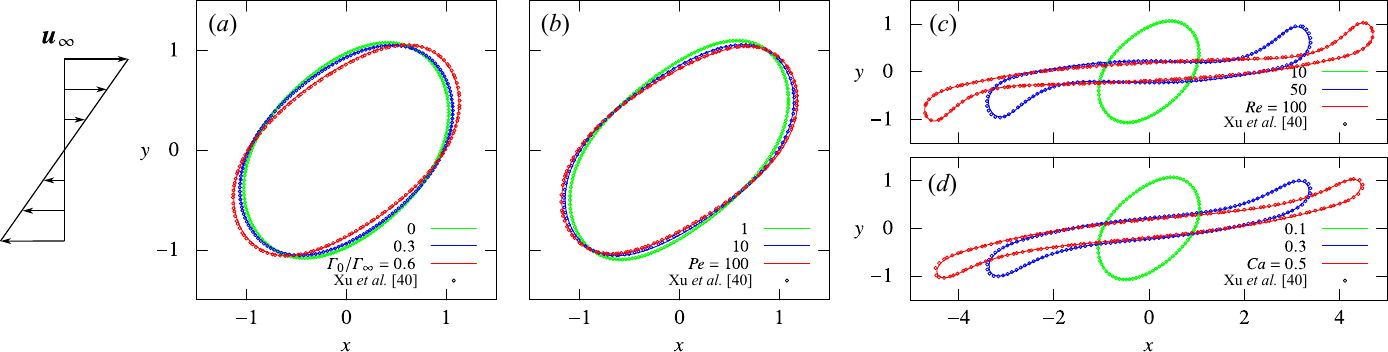}
    \vspace{-5mm}
  \caption{Deformation of a 2D surfactant-laden droplet in a shear flow (illustrated by arrows at the very left). Panels (\textit{a})-(\textit{d}) display the final shape of the droplet corresponding to cases A-D in Table~\ref{tab:shear_flow_2D}. These panels illustrate the effects of $\mathit{\Gamma}_0/\mathit{\Gamma}_\infty$, $\mathit{Pe}$, $\mathit{Re}$, and $\mathit{Ca}$, respectively. Solid lines denote results from the present study, while circles represent reference data from \cite{xu_level-set_2012}.}
  \label{fig:shear_flow_2D}
\end{figure}

Having validated the individual numerical components involved in the simulation of the dynamics of interfacial flows in the presence of surfactants, we are now in position to couple these modules with the Navier-Stokes equations to solve the complete system described by Eqs.~(\ref{eq:ns1})-(\ref{eq:sur-trans}). We start with the deformation of a 2D surfactant-laden droplet (phase 1) undergoing deformation in a linear shear flow.
The computational domain is a $[-5R,5R]\times[-2R,2R]$ rectangle. Initially, the droplet is circular with radius $R$, and the prescribed surfactant concentration is $\mathit{\Gamma}_0$. A linear shear flow is imposed at the domain boundaries in the form $\bm{u} = \gamma y\bm{e}_x$, where $\gamma = \mathrm{d}|\bm{u}|/\mathrm{d}y$ is the shear rate in the bulk phase (phase 2) (see the very left panel in Fig.~\ref{fig:shear_flow_2D}). The nonlinear EOS \eqref{eq:eos} is employed to characterize the relationship between the interfacial tension and the surfactant concentration. This test case was previously used in the validation of the diffuse interface method proposed by Xu \textit{et al.} \cite{xu_level-set_2012}. Considering $R$ and $\gamma^{-1}$ as the characteristic length and time scales, respectively, the relevant dimensionless governing parameters are
\begin{equation}\label{eq:47}
  \rho_r = \frac{\rho_2}{\rho_1},\quad \mu_r = \frac{\mu_2}{\mu_1},\quad \mathit{Re} = \frac{\rho_1\gamma R^2}{\mu_1}, \quad \mathit{Ca} = \frac{\mu_1\gamma R}{\sigma_0}, \quad \mathit{Pe}=\frac{\gamma R^2}{D_s},\quad E = \frac{\mathcal{R}T\mathit{\Gamma}_\infty}{\sigma_0},\quad\frac{\mathit{\Gamma}_0}{\mathit{\Gamma}_\infty}\,,
\end{equation}
where $\mathit{Re}$, $\mathit{Ca}$, $\mathit{Pe}$ and $E$ represent the Reynolds, capillary, P\'{e}clet and elasticity numbers, respectively.
\begin{table}[]
  \centering
  \caption{Physical and numerical parameters used to simulate the deformation of a 2D surfactant-laden droplet in a shear flow. Cases A-D investigate effects of the initial concentration $\mathit{\Gamma}_0/\mathit{\Gamma}_\infty$, P\'{e}clet number ($\mathit{Pe}$), Reynolds number ($\mathit{Re}$), and capillary number ($\mathit{Ca}$), respectively. Case E, which employs a large density ratio, is used for convergence analysis.}
  \label{tab:shear_flow_2D}
  \begin{tabular}{@{}llllllllll@{}}
    \toprule
    case & $\rho_r$ & $\mu_r$ & $\mathit{Re}$ & $\mathit{Ca}$   & $\mathit{Pe}$ & $\mathit{\Gamma}_0/\mathit{\Gamma}_\infty$    & $E$   & $R/\Delta_{\mathrm{min}}$ & $t_f$ \\ \midrule
    A    & 1               & 1             & 10            & 0.1             & 10            & $[0,0.3,0.6]$ & 0.2 & 25.6                      & 9     \\
    B    & 1               & 1             & 10            & 0.1             & $[1,10,100]$  & 0.6           & 0.2 & 25.6                      & 8     \\
    C    & 1               & 1             & $[10,50,100]$ & 0.1             & 10            & 0.3           & 0.2 & 25.6                      & 8     \\
    D    & 1               & 1             & 10            & $[0.1,0.3,0.5]$ & 10            & 0.3           & 0.2 & 25.6                      & 7     \\
    E    & $1/700$         & 0.5           & 10            & 0.7             & 10            & 0.1           & 0.2 & $[12.8,25.6,51.2]$        & 4     \\ \bottomrule
  \end{tabular}
\end{table}

\begin{figure}
  \centering
  \includegraphics[width=\textwidth]{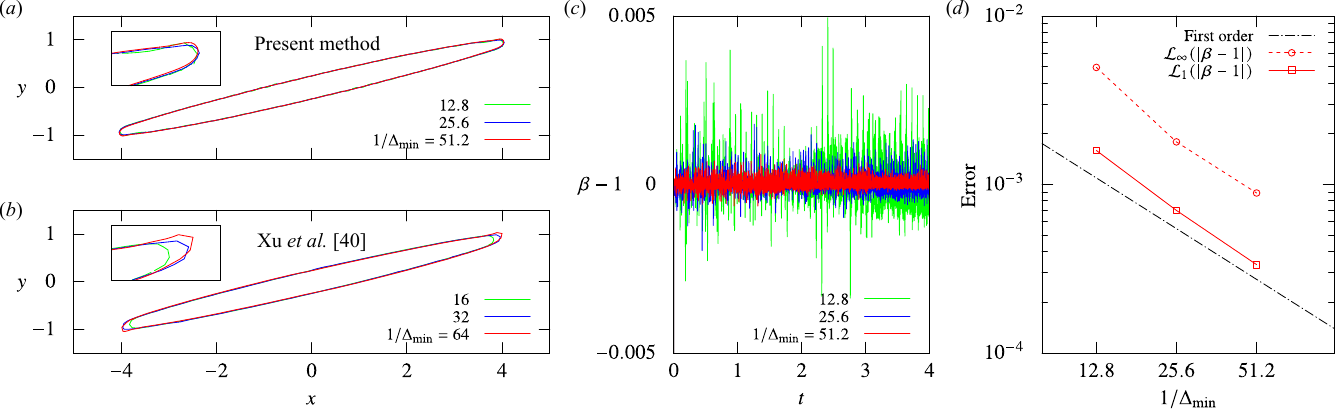}
  \vspace{-5mm}
  \caption{Convergence analysis for case E in Table~\ref{tab:shear_flow_2D}. Panels (\textit{a}) and (\textit{b}) compare the final droplet interfaces obtained using the present method and the reference method from \cite{xu_level-set_2012}, with insets highlighting the tip regions. Panel (\textit{c}) presents the evolution of the rescaling factor $\beta$ for surfactant mass conservation at different resolutions, while panel (\textit{d}) shows the convergence of the $\mathcal{L}_1$ and $\mathcal{L}_\infty$ norms of the deviation $|\beta-1|$ as the mesh is refined.}
  \label{fig:shear_flow_2D_cov}
\end{figure}
Effects of the various physical parameters are investigated through cases A-D, corresponding to different values of $\mathit{\Gamma}_0/\mathit{\Gamma}_\infty$, $\mathit{Pe}$, $\mathit{Re}$, and $\mathit{Ca}$, as summarized in Tab.~\ref{tab:shear_flow_2D}. For these cases, an adaptive grid is employed, achieving a maximum spatial resolution $R/\Delta_{\mathrm{min}}=25.6$ near the interface. $c_{\mathrm{CFL}}$ is set to $0.5$ in all cases. Figure \ref{fig:shear_flow_2D} presents the final droplet shapes, with solid lines denoting present results and circles referring to the reference data of \cite{xu_level-set_2012}.
The presence of surfactants reduces the droplet's interfacial tension while simultaneously accumulating at the droplet tips due to the ambient shear flow, resulting in a minimum interfacial tension at these locations. Consequently, as shown in Fig.~\ref{fig:shear_flow_2D}(\textit{a}), increasing the initial surfactant concentration $\mathit{\Gamma}_0$ leads to greater droplet deformation. In Fig.~\ref{fig:shear_flow_2D}(\textit{b}), a lower $\mathit{Pe}$ enhances surfactant diffusion, which mitigates tip accumulation and consequently limits droplet deformation. Fig.~\ref{fig:shear_flow_2D}(\textit{c}) illustrates the fact that a higher $\mathit{Re}$ amplifies advection effects, leading to an increased deformation. Similarly, in Fig.~\ref{fig:shear_flow_2D}(\textit{d}), a larger $\mathit{Ca}$ reduces the relative influence of interfacial tension compared to viscous forces, further promoting droplet deformation. The results exhibit excellent agreement with the reference data, validating the accuracy and effectiveness of the present method.

Finally, we assess the grid dependence of the method by keeping the physical parameters and $c_{\mathrm{CFL}}$ unchanged while varying the spatial resolution (case E in Tab.~\ref{tab:shear_flow_2D}). Figures \ref{fig:shear_flow_2D_cov}(\textit{a}-\textit{b}) compare the obtained final droplet shapes with those from the reference study \cite{xu_level-set_2012}, with insets providing magnified views of the tip regions. The nearly indistinguishable overlap of interfaces in the insets of Fig.~\ref{fig:shear_flow_2D_cov}(\textit{a}) highlights the much faster convergence of the sharp interface approach as compared with the diffuse interface method. Furthermore, as discussed in \S~\ref{sec:update-surf-conc}, a rescaling step with a correction factor $\beta$ is applied at the end of the advection step to restore the conservation of the surfactant mass altered by the PLIC scheme. The evolution of the deviation $\beta - 1$ for different resolutions is presented in Fig.~\ref{fig:shear_flow_2D_cov}(\textit{c}) and its convergence is detailed in Fig.~\ref{fig:shear_flow_2D_cov}(\textit{d}). These results confirm that $|\beta - 1|$ remains very small, of the order of $10^{-3}$ throughout the simulation, and its $\mathcal{L}_1$ and $\mathcal{L}_\infty$ norms exhibit first-order convergence with mesh refinement.

\subsection{Deformation of a 3D surfactant-laden drop in an extensional flow} \label{sec:dropl-deform-an}

\begin{figure}
  \centering
  \includegraphics[width=\textwidth]{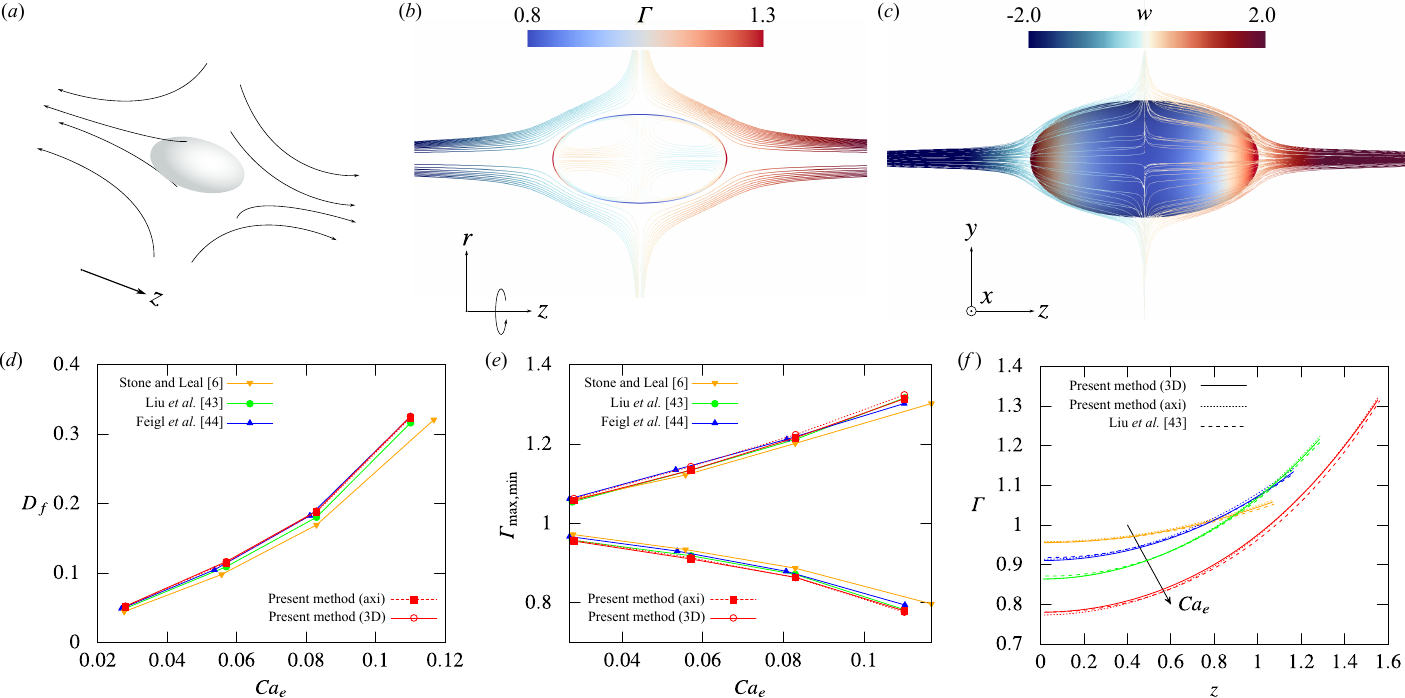}
  \vspace{-5mm}
  \caption{Deformation of a droplet in a 3D extensional flow. (\textit{a}) Sketch of the physical problem; (\textit{b}) and (\textit{c}) droplet interface (colored by surfactant concentration), and streamlines (colored by the axial velocity, $w$), at the final stage for $\mathit{Ca}_e=0.11$ in the axisymmetric and 3D cases, respectively. Both plots share the same color scales. (\textit{d}) Deformation factor, $D_f$; (\textit{e}) maximum and minimum surfactant concentrations, $\mathit{\Gamma}_{\mathrm{max}}$ and $\mathit{\Gamma}_{\mathrm{min}}$, as function of $\mathit{Ca}_e$ at the final stage; (\textit{f}) surfactant concentration distribution along the axial direction for different $\mathit{Ca}_e$.
  }
  \label{fig:extensional}
\end{figure}

We now validate the fully coupled 3D method for Eqs.~(\ref{eq:ns1}) - (\ref{eq:sur-trans}) using the test case of Liu \textit{et al.} \cite{liu_hybrid_2018} in which a surfactant-laden spherical droplet with radius $R=1$ deforms under an extensional flow, as illustrated by Fig.~\ref{fig:extensional}(\textit{a}). Initially, the droplet is laden with a uniform surfactant concentration, and the problem is inherently axisymmetric. This is why we perform simulations in both axisymmetric and fully 3D geometries. The far-field velocity is prescribed as
\begin{equation}\label{eq:48}
  \bm{u}_\infty(\bm{x}) = \gamma
  \begin{pmatrix}
    -\frac{1}{2}&0&0\\
    0&-\frac{1}{2}&0\\
    0&0&1
  \end{pmatrix}
  \cdot
  \bm{x}\,,
\end{equation}
where $\gamma$ denotes the strain rate.
The droplet (phase 1) and the carrying liquid (phase 2) have identical density and viscosity and we assume a linear equation of state (EOS), $\sigma = \sigma_0 - \mathcal{R}T\mathit{\Gamma}$.
Using $R$ and $\gamma^{-1}$ as the characteristic length and time scales, the droplet dynamics are governed by the same dimensionless numbers as in Eq.~(\ref{eq:47}). Following Liu \textit{et al.} \cite{liu_hybrid_2018}, we set $E=0.1$, $\mathit{Pe} = 10\mathit{Ca}_e$, $\mathit{Re}=0.05$, and $\mathit{\Gamma}_0/\mathit{\Gamma}_\infty = 1$. The effective capillary number is defined as $\mathit{Ca}_e = \mathit{Ca}/(1-E)$, with values $\mathit{Ca}_e = 0.028, 0.057, 0.083$, and $0.11$. In the axisymmetric case, the computational domain extends $5R$ from the droplet center in both the axial and radial directions, with the far-field velocity imposed at all boundaries. In the 3D simulation, we exploit the existing flow symmetries and simulate only one-eighth of the domain, with the droplet centered at $(x,y,z)=(0,0,0)$ within a $[0,5R]\times[0,5R]\times[0,5R]$ domain. The far-field velocity is applied on planes $x=5R$, $y=5R$, and $z=5R$, while symmetry conditions are enforced on planes $x=0$, $y=0$, and $z=0$. $c_{\mathrm{CFL}}$ is still set to 0.5, and the AMR procedure refines the spatial resolution up to $1/\Delta_{\mathrm{min}} = 25.6$ near the interface.

Numerical results are summarized in Fig.~\ref{fig:extensional}. Panels (\textit{b}) and (\textit{c}) illustrate the steady-state droplet shape, surfactant concentration, and streamlines obtained with $\mathit{Ca}_e=0.11$ for the axisymmetric and 3D cases, respectively. The extensional flow stretches the droplet along the $z$ axis and surfactants accumulate at the droplet ends, leading to higher concentrations at the poles and lower concentrations at the equator. The droplet deformation is quantified using the familiar deformation factor $D_f = (R_z - R_x)/(R_z + R_x)$, with $R_z$ and $R_x$ being the half-lengths of the major and minor axes, respectively. With no surprise, Fig. \ref{fig:extensional}(\textit{d}) shows that $D_f$ increases with $\mathit{Ca}_e$, owing to the weakening restoring interfacial tension force. Since the P\'eclet number is set to $\mathit{Pe} = 10\mathit{Ca}_e$,  increasing $\mathit{Ca}_e$ increases $\mathit{Pe}$, which in turn reduces the net effect of surfactant diffusion, leading to a more pronounced surfactant accumulation at the droplet ends, as shown in Fig.~\ref{fig:extensional}(\textit{e}). We compare present predictions with those from \cite{stone_effects_1990,feigl_simulation_2007,liu_hybrid_2018}. Excellent agreement is observed with the results of \cite{feigl_simulation_2007} and \cite{liu_hybrid_2018}, confirming the accuracy of the fully coupled method in both axisymmetric and 3D geometries. Differences may be noticed with predictions of Stone \textit{et al.} \cite{stone_effects_1990}, likely due to their use of the Stokes flow approximation. Further validation is provided in Fig.~\ref{fig:extensional}(\textit{f}), which presents the steady-state surfactant concentration distribution along the droplet axis for different $\mathit{Ca}_e$. As expected, larger $\mathit{Ca}_e$ results in more elongated droplets and stronger surfactant accumulation at the poles. Both the axisymmetric and 3D results closely agree with those of \cite{liu_hybrid_2018}.

\section{Rise of a three-dimensional surfactant-laden bubble close to a vertical wall} \label{sec:numerical-examples}

\begin{figure}[h]
  \centering
  \includegraphics[width=\textwidth]{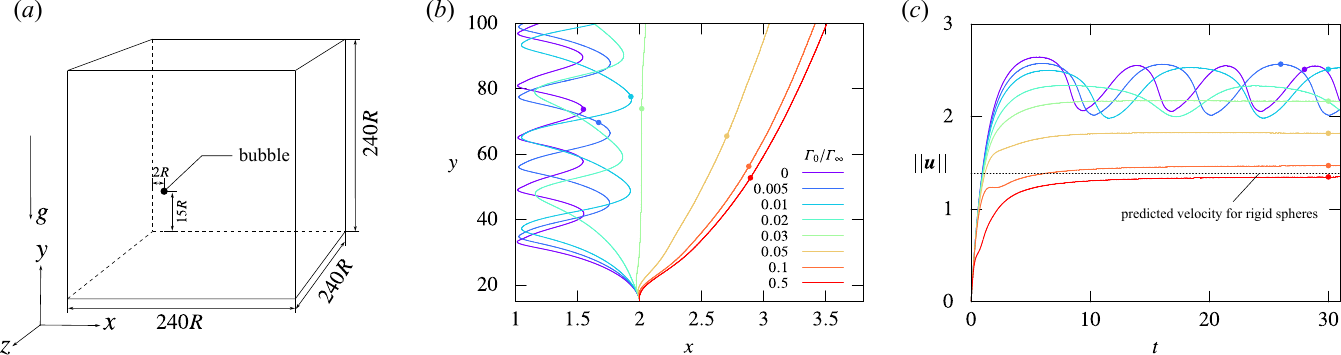}
    \vspace{-5mm}
  \caption{Interaction of a surfactant-laden rising bubble with a vertical wall. (\textit{a}) Sketch of the physical configuration (not to scale); (\textit{b}) path of bubbles with different initial surfactant concentrations, $\mathit{\Gamma}_0/\mathit{\Gamma}_\infty$; (\textit{c}) evolution of the norm of the bubble velocity for different $\mathit{\Gamma}_0/\mathit{\Gamma}_\infty$. The black dotted line in (\textit{c}) denotes the terminal velocity of a rigid sphere predicted by the Schiller-Neumann drag correlation \cite{clift1978bubbles}. Solid markers in (\textit{b}-\textit{c}) indicate moments corresponding to the snapshots in Fig.~\ref{fig:shi-field}.}
  \label{fig:shi}
\end{figure}

\begin{figure}[h]
  \centering
  \includegraphics[width=.9\textwidth]{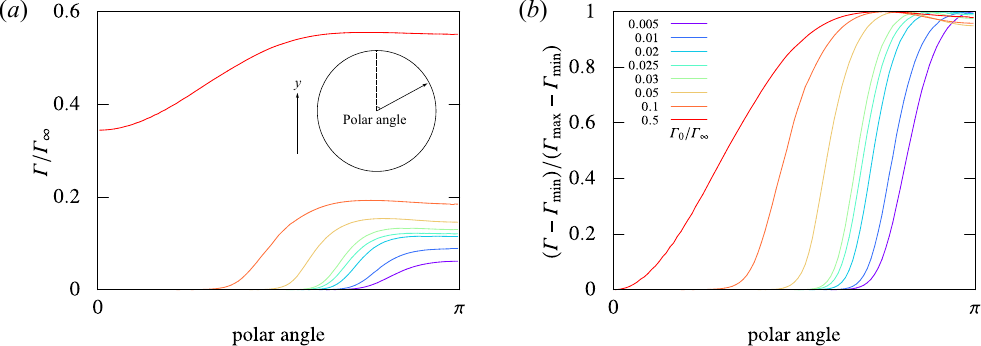}
  \caption{(\textit{a}) Surfactant distribution along the bubble interface at final state, for different $\mathit{\Gamma}_0/\mathit{\Gamma}_\infty$ (the polar angle is defined in the inset); (\textit{b}) same as (\textit{a}), but with the surfactant concentration variations normalized by the difference between the maximum and minimum concentrations, $\mathit{\Gamma}_{\mathrm{max}}$ and $\mathit{\Gamma}_{\mathrm{min}}$.}
  \label{fig:shi-profile}
\end{figure}

\begin{figure}[h]
  \centering
  \includegraphics[width=\textwidth]{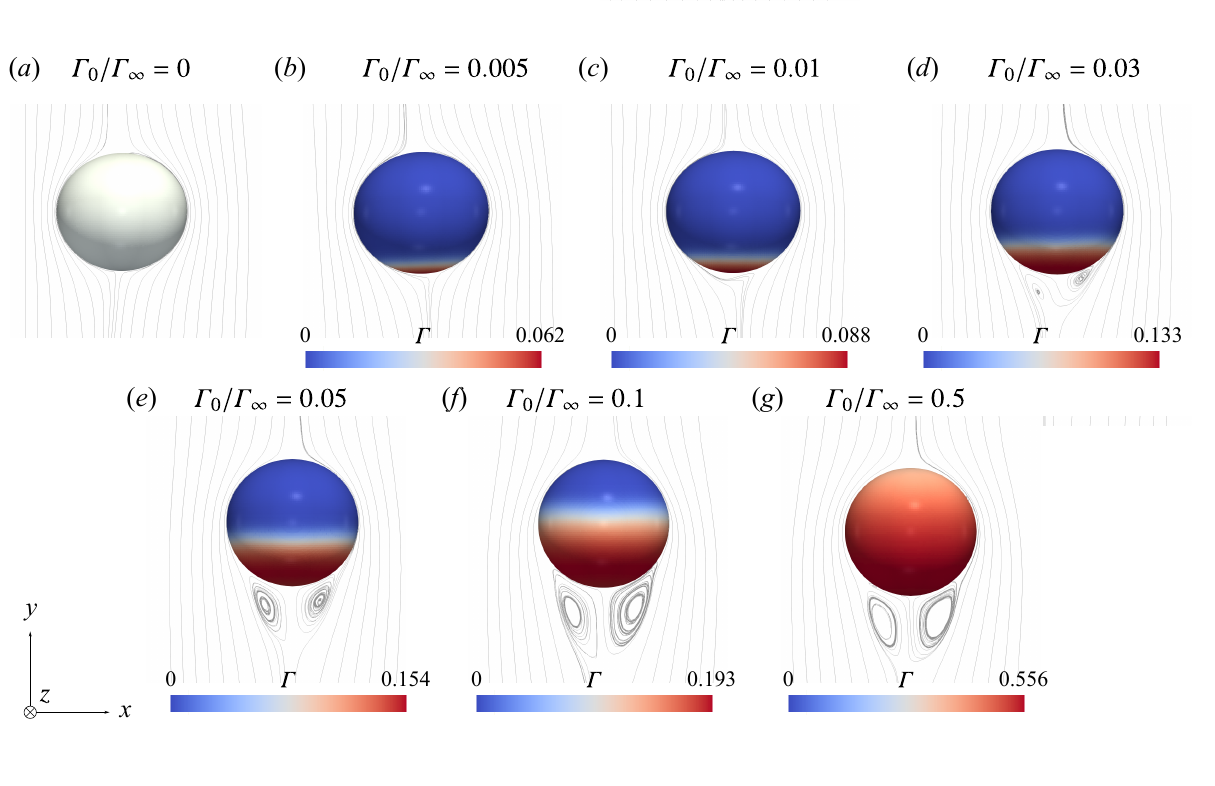}
    \vspace{-10mm}
  \caption{Surfactant distribution along the bubble interface and streamlines in the symmetry plane $z=0$ in the final state, for different initial concentrations $\mathit{\Gamma}_0/\mathit{\Gamma}_\infty$. Panels (\textit{a}-\textit{g}) correspond to $\mathit{\Gamma}_0/\mathit{\Gamma}_\infty = 0,0.005,0.01,0.03,0.05,0.1,0.5$, respectively. Streamlines are shown in the bubble's reference frame.}
  \label{fig:shi-field}
\end{figure}

As a final complex test, we investigate the interaction of a 3D surfactant-laden rising bubble with a vertical flat wall, to demonstrate the capability of the proposed method in predicting subtle physical behaviors in an unsteady, long-duration problem. The setup is based on an experimental study by Takemura and Magnaudet \cite{takemura_transverse_2003}, which examined the transverse migration of both clean and fully contaminated nearly-spherical bubbles rising near a vertical wall in a quiescent liquid.


The computational set-up, shown in Fig.~\ref{fig:shi}(\textit{a}), makes use of a cubic domain $[0,240R]\times[0,240R]\times[0,240R]$, with $R$ denoting the initial bubble radius. A no-slip boundary condition is imposed on the left ($x=0$), top ($y=240R$), and bottom ($y=0$) walls; a free-slip condition is applied on the other boundaries. The left wall is assumed hydrophilic with a phase indicator $c=1$ throughout the simulation ($c>0$ in a cell where the liquid phase (phase 1) is present). Gravity $\bm{g} = -g\bm{e}_y$ is acting downwards. The center of the initially stationary spherical bubble is placed at $(x, y, z) = (2R, 15R, 0)$, such that the gap between the bubble and wall is one radius at $t=0$. For numerical efficiency, only half of the flow field (say the half-plane $z\leq0$) is simulated, since preliminary numerical studies for non-contaminated bubbles \cite{shi_lateral_2024} showed that the flow remains symmetric with respect to the $z=0$ plane within the considered range of physical parameters. The nonlinear EOS (\ref{eq:eos}) is used to model the interfacial tension dependence on surfactant concentration. The AMR procedure is applied with a spatial resolution up to $R/\Delta_{\mathrm{min}} \approx 34$ near the interface. The bubble position, rise speed, and time are normalized by $R$, $(gR)^{1/2}$, and $(R/g)^{1/2}$, respectively. The bubble and flow dynamics are governed by the set of dimensionless parameters
\begin{equation}
  \rho_r = \frac{\rho_2}{\rho_1},\quad
  \mu_r = \frac{\mu_2}{\mu_1},\quad
  \mathit{Ga} = \frac{\rho_1g^{1/2}R^{3/2}}{\mu_1},\quad
  \mathit{Bo} = \frac{\rho_1gR^2}{\sigma_0},\quad
    \mathit{Pe} = \frac{g^{1/2}R^{3/2}}{D_s},\quad
  E = \frac{\mathcal{R}T\mathit{\Gamma}_\infty}{\sigma_0},\quad
  \frac{\mathit{\Gamma}_0}{\mathit{\Gamma}_\infty}\,.
\end{equation}
To mimic realistic conditions, the density and viscosity ratios are set to $\rho_r = 10^{-3}$ and $\mu_r = 10^{-2}$, respectively. We also set $E = 0.1$ and $\mathit{Pe} = 10$.

For a clean, deformable bubble, a recent numerical study by Shi \textit{et al.} \cite{shi_lateral_2024} identified three distinct regimes of bubble-wall interaction. At low $\mathit{Ga}$ and high $\mathit{Bo}$, the bubble deforms significantly and is repelled from the wall due to the wake-wall interaction. At high $\mathit{Ga}$ and low $\mathit{Bo}$, the bubble remains nearly spherical and undergoes periodic near-wall bouncing, driven by an alternating dominance of attractive irrotational (Bernoulli-like) effects and wake-wall interactions. At moderate $\mathit{Ga}$ and $\mathit{Bo}$, the bubble initially approaches the wall before stabilizing at an equilibrium lateral position, where the transverse forces produced by the two mechanisms are in balance. Takemura and Magnaudet \cite{takemura_transverse_2003} also highlighted that surfactants at the interface dramatically alter the bubble-wall interaction by increasing the amount of vorticity produced at the bubble surface, which translates into stronger wake effects, hence a stronger repelling force. Specifically, they found that for rise Reynolds numbers exceeding 35, the overall transverse force directs a clean bubble towards the wall, whereas contaminated bubbles are repelled from the wall whatever the Reynolds number.

To systematically examine surfactant effects, we define an effective Bond number $\mathit{Bo}_e = \frac{\mathit{Bo}}{1+E \ln(1-\mathit{\Gamma}_0/\mathit{\Gamma}_\infty)}$ which is kept constant across cases to facilitate comparison. We select $(\mathit{Bo}_e, \mathit{Ga}) = (0.073, 21.9)$ while varying $\mathit{\Gamma}_0/\mathit{\Gamma}_\infty$ in the range $[0,0.5]$. The reason for selecting this specific $(\mathit{Bo}_e, \mathit{Ga})$ pair is that this case was considered in great detail in \cite{shi_lateral_2024} for a perfectly clean bubble ($\mathit{\Gamma}_0=0$). Figure \ref{fig:shi}(\textit{b}) presents bubble trajectories for different surfactant concentrations. The clean bubble ($\mathit{\Gamma}_0/\mathit{\Gamma}_\infty = 0$) initially moves toward the wall and undergoes periodic bouncing with a fixed amplitude and frequency, consistent with the findings of \cite{shi_lateral_2024}. For $\mathit{\Gamma}_0/\mathit{\Gamma}_\infty > 0$, a transition from periodic bouncing to migration away from the wall is seen to take place for $\mathit{\Gamma}_0/\mathit{\Gamma}_\infty\approx0.03$. At lower concentrations ($\mathit{\Gamma}_0/\mathit{\Gamma}_\infty = 0.005, 0.01, 0.02$), the periodic bouncing observed with a clean bubble persists. However, the larger the initial concentration, the larger the amplitude and the lower the frequency of the bouncing motion. At higher concentrations ($\mathit{\Gamma}_0/\mathit{\Gamma}_\infty = 0.03, 0.05, 0.1, 0.5$), the bubble is no longer deviated towards the wall, and the repelling effect becomes more pronounced with increasing surfactant concentration. Figure \ref{fig:shi}(\textit{c}) shows the corresponding evolution of the bubble velocity, in which periodic oscillations are observed in the bouncing cases. The terminal velocity (estimated by averaging the instantaneous velocity over one bouncing period) is seen to decrease with increasing concentration. In particular, the terminal velocity found for $\mathit{\Gamma}_0/\mathit{\Gamma}_\infty = 0.5$ approaches that of a freely-rising rigid sphere (dotted line), indicating that the outer fluid essentially obeys a no-slip condition at the interface.
The underlying mechanism is well understood: surfactants are advected from the bubble front to the rear, generating a negative interfacial tension gradient ($\partial\sigma/\partial\theta<0$) that induces a Marangoni stress from rear to front, since $d\sigma/d\mathit{\Gamma}<0$. This Marangoni stress is balanced by a shear stress directed towards the bubble rear which enhances the drag, therefore reducing the bubble rise speed. This shear stress also results in a larger vorticity around the bubble, hence a stronger wake, enhancing the repelling bubble-wall interaction.

Figure \ref{fig:shi-profile}(\textit{a}) presents the steady-state surfactant distribution $\mathit{\Gamma}(\theta)$ along the bubble surface for different $\mathit{\Gamma}_0/\mathit{\Gamma}_\infty$, while Fig. \ref{fig:shi-profile}(\textit{b}) displays the normalized distribution $(\mathit{\Gamma}(\theta)-\mathit{\Gamma}_\mathrm{min})/(\mathit{\Gamma}_\mathrm{max}-\mathit{\Gamma}_\mathrm{min})$. 
For $\mathit{\Gamma}_0/\mathit{\Gamma}_\infty \leq 0.1$, the front part of the interface remains free of surfactant up to a critical angle $\theta=\theta_\mathrm{inf}$. This critical angle reduces as $\mathit{\Gamma}_0$ increases, a behavior consistent with the well-known stagnant cap model \cite{palaparthi_theory_2006}. In contrast, the surfactant covers the entire interface for $\mathit{\Gamma}_0/\mathit{\Gamma}_\infty = 0.5$. This is because $\mathit{\Gamma}_0$ is too close to the maximum packing concentration $\mathit{\Gamma}_\infty$ for a large concentration gradient around an intermediate $\theta$ to be possible without reaching local concentrations exceeding $\mathit{\Gamma}_\infty$. Figure \ref{fig:shi-field} displays the wake structure and streamlines for various $\mathit{\Gamma}_0/\mathit{\Gamma}_\infty$, with colors used to represent iso-values of the surfactant distribution on the bubble half-surface facing the wall. As $\mathit{\Gamma}_0$ increases, the wake becomes more pronounced. A closed recirculating region is observed when $\mathit{\Gamma}_0\gtrsim0.03$. The size of this eddy increases with the surfactant concentration up to $\mathit{\Gamma}_0/\mathit{\Gamma}_\infty=0.1$, making the bubble behave increasingly as a rigid sphere. In contrast, the eddy is seen to be slightly smaller for $\mathit{\Gamma}_0/\mathit{\Gamma}_\infty = 0.5$ as compared with $\mathit{\Gamma}_0/\mathit{\Gamma}_\infty = 0.1$, owing to the somewhat lower bubble rise speed (hence, Reynolds number) at the highest concentration.
\section{Summary}\label{sec:conclusion}

We have developed a three-dimensional sharp-interface numerical approach for simulating interfacial flows with insoluble surfactants within the geometrical VOF framework of the \textit{Basilisk} open software. This method is consistent with the zero-thickness property of the interface, ensures accurate computation of surfactant concentration, and accommodates arbitrary geometrical changes of the interface. The surfactant mass is stored sharply at the interface reconstructed through the geometrical VOF method. The advection term of the surfactant transport equation is discretized implicitly alongside the geometrical advection imposed by the VOF equation, eliminating numerical inconsistencies between the actual and computed interface areas. The surface diffusion term is discretized in a conservative and sharp manner within a finite-volume framework along the reconstructed interface, preventing artificial diffusion normal to the interface.

Compared to existing {Eulerian}-based methods, the present approach introduces several key improvements. Unlike the widely used diffuse approaches, which artificially smear the interfacial region, our technique maintains the interface sharpness, thereby improving both physical representativity and numerical accuracy. Moreover, the earlier attempt of \cite{james_surfactant-conserving_2004} to treat surfactant transport in the framework of a {VOF}-based method explicitly solves an advection-stretching equation for the interface area, leading to inconsistencies with the estimate of the reconstructed interface area implicitly dictated by the VOF advection scheme. In contrast, our method explicitly incorporates the area change of the reconstructed interface, ensuring accurate surfactant advection in a sharp manner. This improvement allows consistent long-term simulations that cannot be achieved with the approach of \cite{james_surfactant-conserving_2004}. 

We validated the accuracy and robustness of the proposed method through a thorough set of test cases, demonstrating improved accuracy and faster convergence compared with existing diffuse interface methods. We further applied this method to investigate the interaction of a surfactant-laden rising bubble with a vertical wall, revealing a transition from periodic near-wall bouncing to migration away from the wall due to the extra vorticity generation at the bubble surface inherent to the presence of surfactants.

In real-world applications, surfactants most often migrate from the bulk to the interface and \textit{vice versa}, owing to adsorption and desorption processes. Numerically capturing the bulk-interface surfactant exchange while maintaining interface sharpness and mass conservation presents significant additional challenges. The extension of the numerical approach presented here to accurately simulate the dynamics of interfacial flows in the presence of soluble surfactants will be the subject of a forthcoming paper.

\section{Acknowledgement}

The authors gratefully acknowledge the support of the National Key R$\&$D Program of China under grants number 2023YFA1011000 and 2022YFE03130000, that of the NSFC under grants numbers 12222208, 124B2046 and 12472256 and that of the ``the Fundamental Research Funds for the Central Universities'' under grants number xzy022024050.

\appendix

\section{Mass flux calculation in axisymmetric configurations}\label{sec:mass-flux-calc}
\begin{figure}
  \centering
  \includegraphics[width=.35\textwidth]{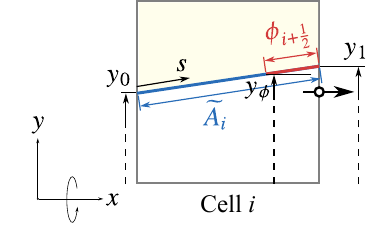}
  \caption{Illustration of the mass flux computation at the right cell face in an axisymmetric configuration. The symmetry axis lies along the $x$ coordinate and $y$ is the radial coordinate.}
  \label{fig:mass-flux-axi}
\end{figure}

In this appendix, we extend the 2D technique detailed in \S~\ref{sec:mass-flux-estimation} to axisymmetric configurations. The core concept remains the same, but the mathematical expressions are more complex.
We focus on the mass flux computation at the right cell face, where $x$ is the symmetry axis and $y$ is the radial direction (see Fig.~\ref{fig:mass-flux-axi}). Since $y$ may vary along the reconstructed interface, we first assume that, along the arc-length abscissa $s$, the surfactant concentration varies as
\begin{equation}\label{eq:a1}
  \mathit{\Gamma}(s) = s(\nabla_s\mathit{\Gamma})_i + \mathit{\Gamma}_c\,,
\end{equation}
where $\mathit{\Gamma}_c$ is a constant to be determined from mass conservation. The linear reconstructed interface can be parameterized as $x = ky+x_0$ with slope $k$ and intercept $x_0$. Thus $s = \sqrt{1+k^2}(y-y_0)$ and $\mathrm{d}s = \sqrt{1+k^2}~\mathrm{d}y$. Then Eq.~(\ref{eq:a1}) can be rewritten as
\begin{equation}\label{eq:a2}
  \mathit{\Gamma}(y) = (\nabla_s\mathit{\Gamma})_i\sqrt{1+k^2}(y-y_0) + \mathit{\Gamma}_c.
\end{equation}
Using mass conservation of surfactant in cell $i$, the averaged concentration is expressed as:
\begin{equation}\label{eq:a3}
  \mathit{\Gamma}_i = \frac{\int_{y_0}^{y_1} 2\pi y \sqrt{1+k^2} \mathit{\Gamma}(y)~\mathrm{d}y } {\int_{y_0}^{y_1} 2\pi y \sqrt{1+k^2}~\mathrm{d}y }.
\end{equation}
Making use of (\ref{eq:a2}) then yields the relation
\begin{equation}\label{eq:a4}
  \mathit{\Gamma}_c = \mathit{\Gamma}_i - (\nabla_s\mathit{\Gamma})_i \sqrt{1+k^2}\left[\frac{2(y_1^3-y_0^3)}{3(y_1^2-y_0^2)}-y_0\right].
\end{equation}
Finally, if the area flux $\phi_{i + \frac{1}{2}}$ is positive, integration is performed along the right side of the interface segment, as indicated by the red portion in Fig.~\ref{fig:mass-flux-axi}. The mass flux is then obtained as
\begin{equation}\label{eq:a5}
  \psi_{i+\frac12} = \int_{y_{\phi}}^{y_1} 2\pi\sqrt{1+k^2}\left[(\nabla_s\mathit{\Gamma})_i\sqrt{1+k^2}(y^2-y_0y)+\mathit{\Gamma}_c y\right]~\mathrm{d}y\,,
\end{equation}
where $y_\phi$ is obtained from
\begin{equation}
  \frac{\phi_{i+\frac12}}{\widetilde A_i} =\frac{\int_{y_\phi}^{y_1}2\pi y~dy}{\int_{y_0}^{y_1}2\pi y ~dy} =
  \frac{y_1^2-y^{2}_{\phi}}{y_1^2-y_0^2} \Rightarrow y_\phi = \sqrt{y_1^2 - \frac{\phi_{i+\frac12}}{\widetilde A_i}(y_1^2-y_0^2)}\,.
\end{equation}
Extensions to configurations with negative $\phi_{i + \frac{1}{2}}$ and to the mass flux in the radial direction are straightforward.

\section{Mass flux calculation in 3D configurations}\label{sec:mass-flux-calc-1}

\begin{figure}[h]
  \centering
  \includegraphics[width=\textwidth]{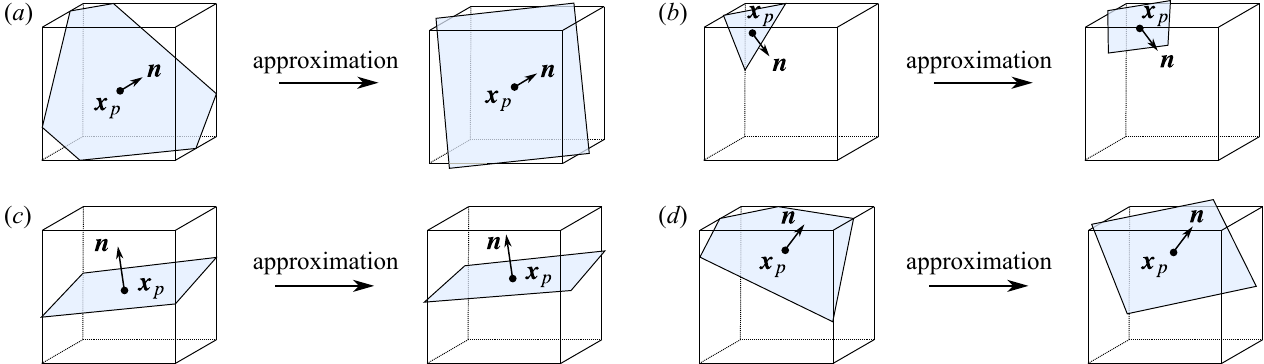}
    \vspace{-3mm}
  \caption{Illustration of the approximation applied to the 3D VOF interface to simplify the area flux computation. All local 3D VOF interfaces are approximated by squares having the same area, normal vector $\bm{n}$, and centroid $\bm{x}_p$ as the original VOF interface, and have one side perpendicular to the advection direction. Panels (\textit{a}-\textit{d}) depicts this approximation for the four types of possible configurations of the 3D interface within a cubic cell.}
  \label{fig:mass-flux-3d-appro}
\end{figure}

\begin{figure}[h]
  \centering
  \includegraphics[width=.65\textwidth]{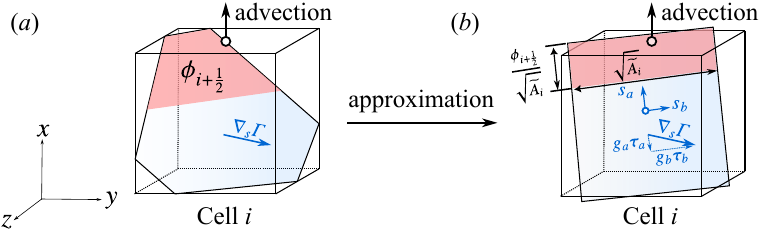}
  \caption{Illustration of the area flux computation in the positive $x$ axis direction, using the proposed approximation for the 3D VOF interface shape depicted in Fig.~\ref{fig:mass-flux-3d-appro}. (\textit{a}) Before the approximation, obtaining the surfactant mass requires integration over the irregular pink area, based on the surface gradient $\nabla_s\mathit{\Gamma}$, which implies considering numerous specific cases. (\textit{b}) After the approximation is made, integration becomes similar to that in 2D configurations (see Fig.~\ref{fig:area-flux-estimation}(\textit{c})).}
  \label{fig:mass-flux-3d-adv}
\end{figure}

We now extend the 2D technique to 3D configurations.. The core concept remains unchanged. However, the local 3D VOF interface is generally irregular, yielding different contributions from each edge to the total area flux in a given cell. Inferring the surfactant mass from these different contributions (highlighted by the pink area in Fig.~\ref{fig:mass-flux-3d-adv}(\textit{a})) is complex, as numerous specific cases have to be considered.

To simplify the calculation, we note that the piece of interface within a cell may always be approximated by a square satisfying the following conditions (Fig.~\ref{fig:mass-flux-3d-appro}): (i) it has the same area, normal vector $\bm{n}$, and centroid $\bm{x}_p$ as the original piece; (ii) one side of the square is perpendicular to the advection direction. Panels (\textit{a}-\textit{d}) of Fig.~\ref{fig:mass-flux-3d-appro} depict the corresponding approximation for the four possible geometric configurations of the interface within.a cell.

Next, consider the reconstructed interface in Fig.~\ref{fig:mass-flux-3d-appro}(\textit{a}) as an example. When advected in the positive $x$ direction, the original nonuniform area flux is transformed into a uniform flux, as highlighted by the pink region in Fig.~\ref{fig:mass-flux-3d-adv}(\textit{b}). As previously mentioned, this approximation leaves both the area $A_i$ and the area flux $\phi_{i+1/2}$ unchanged. Therefore, the area flux density that leaves the cell $i$ is $\phi_{i+1/2}/\sqrt{A_i}$, and the length of the cell edge crossed by this flux is $\sqrt{A_i}$.

To compute the surfactant mass lying on this piece of interface, we also need to estimate the orthogonal projections of the surface gradient, $\nabla_s\mathit{\Gamma}$. Let $\bm{\tau}_b$ denote the unit vector tangent to the side of the approximating square that is perpendicular to the advection direction, and $\bm{\tau}_a$ the one tangent to the perpendicular side. The surface gradient $\mathit{\Gamma}$ may be expressed as $\nabla_s\mathit{\Gamma} = g_a\bm{\tau}_a + g_b\bm{\tau}_b$, with $g_a = (\nabla_s\mathit{\Gamma})\cdot\bm{\tau}_a$ and $g_b = (\nabla_s\mathit{\Gamma})\cdot\bm{\tau}_b$ (see Fig.~\ref{fig:mass-flux-3d-adv}). Additionally, we introduce a local system of orthogonal coordinates $(s_a,s_b)$  on the surface, with $s_a$ along $\bm{\tau}_a$ and $s_b$ along $\bm{\tau}_b$, and assume the position of the interface centroid $\bm{x}_p$ to be $(0,0)$ in that system.

The mass flux $\psi_{i+1/2}$ (pink region in Fig.~\ref{fig:mass-flux-3d-adv}(\textit{b})) is then estimated as
\begin{align}
  \psi_{i+\frac12} &=  \int_{-\frac12\sqrt{A_i}}^{\frac12\sqrt{A_i}}\int_{\frac12\sqrt{A_i} - \frac{\phi_{i+\frac12}}{\sqrt{A_i}}}^{\frac12\sqrt{A_i}} (\mathit{\Gamma}_i + g_bs_b+ g_a s_a) ~\mathrm{d}s_a~\mathrm{d}s_b \nonumber \\
                   & =\sqrt{A_i}\left[\frac{\mathit{\Gamma}_i\phi_{i+\frac12}}{\sqrt{A_i}} +  \frac12g_a\left(\phi_{i+\frac12} - \frac{\phi_{i+\frac12}^2}{A_i}\right)\right]\,,
\end{align}
with $\mathit{\Gamma}_i$ being the concentration at the interface centroid.

\section{Comparison between options I and II for the concentration calculation}\label{sec:comp-betw-opti}

\begin{figure}
  \centering
  \includegraphics[width=\textwidth]{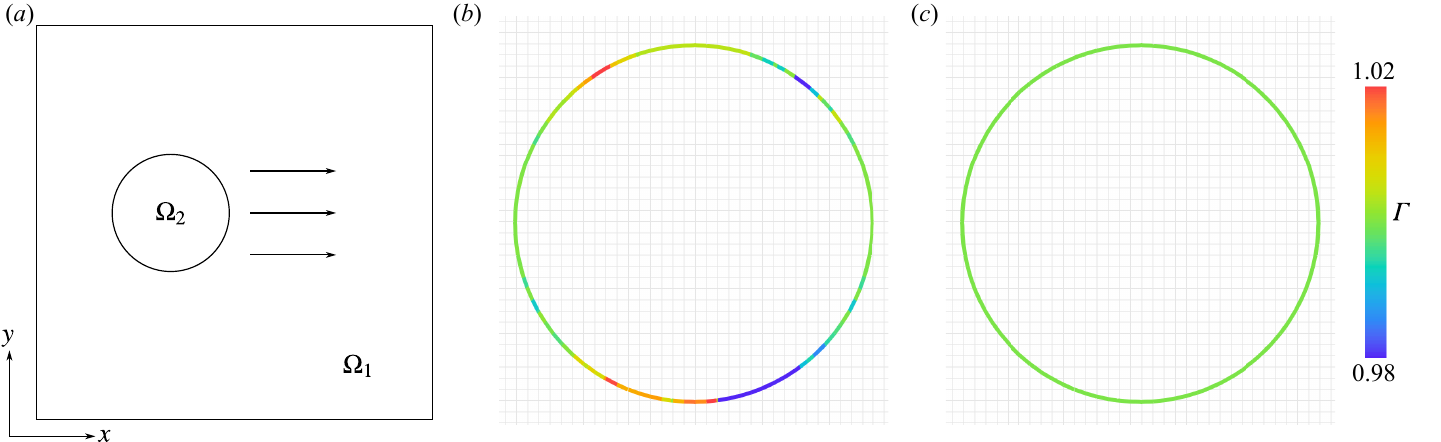}
    \vspace{-3mm}
  \caption{Comparison of options I and II from \S~\ref{sec:update-surf-conc} in calculating the surfactant concentration  in the case of a translating surfactant-laden circular droplet with a uniform concentration $\mathit{\Gamma}(t=0)=1$. (\textit{a}): schematic illustration of the test case (not to scale). (\textit{b}) (resp. (\textit{c})): concentration distribution obtained after 10 time steps with option I (resp. II).}
  \label{fig:two-options}
\end{figure}

To illustrate the rationale for selecting option II over option I, as proposed in \S~\ref{sec:update-surf-conc}, we conduct a simple test. In a two-dimensional domain $[-4,4]\times[-4,4]$, a surfactant-laden circular droplet with an initial unit radius and a uniform concentration $\mathit{\Gamma}(t=0)=1$ is centered at $(-1,0)$. The droplet is subjected to a constant velocity field $(u,v) = (1,0)$, translating along the positive $x$ axis, as shown in Fig.~\ref{fig:two-options}(\textit{a}). A uniform mesh with $1/\Delta = 16$ is used, and $c_{\mathrm{CFL}}$ is set to 0.1. Figures \ref{fig:two-options}(\textit{b}-\textit{c}) show the surfactant concentration distribution after 10 time steps with the two options. When using option I, with the reconstructed interface area $A$ employed to go from the surfactant mass $M$ to the concentration $\mathit{\Gamma}=M/A$, the latter fails to remain constant. The corresponding variations result in a non-negligible numerical Marangoni force even after just 10 time steps. In contrast, option II maintains a constant concentration distribution up to machine accuracy.

\bibliographystyle{elsarticle-num}
\bibliography{mybib}

\end{document}